
\magnification=\magstep1\openup3pt
\parskip=3pt plus 1pt minus .5pt \raggedbottom \topskip=\baselineskip


\def\foot#1#2{\footnote{\spaceskip=0pt#1\spaceskip=0pt}{\openup-1\jot
{\eightpoint\hskip-17pt#2}\tenpoint\openup1\jot}}
\def\footdate#1{\footline{\hss\tenrm--\ #1\ --\hss}}
\def\preprintno#1{\headline={\rightline{\vbox{\halign{\hfil##\cr#1\crcr}}}}}
\def\section#1\par{\vskip18mm\vskip0pt plus\baselineskip\penalty-250\vskip0pt
plus-\baselineskip\medbreak\message{#1}\leftline{\U{\bf#1}}\nobreak\smallskip
\indent}
\def\subsection#1\par{\vskip\baselineskip\medbreak\message{#1}\leftline{\bf#1}
\nobreak\indent}
\def\references#1{\par\vskip18mm\medbreak\message{References}
\leftline{\U{\bf References}}\nobreak\smallskip{\halign{\vtop{\parskip=0pt
\parindent=0pt \hangindent=5mm\strut##\strut}\cr#1\crcr}}}
\def\case #1: #2{\par\medbreak\noindent{\it#1.\enspace}#2}
\def\title#1{\halign{\bold\centerline{##}\hfil\cr#1\crcr}\vskip5mm}
\def\authors#1{\vskip5mm{\halign{\centerline{##}\hfil\cr#1\crcr}}}
\def\address#1{\vskip5mm{\halign{\it\centerline{##}\hfil\cr#1\crcr}}\vskip5mm}
\def\abstract{\vskip15mm{\bf\centerline{Abstract}}\vskip8mm}
\def\endpagefoot#1{\vfill\vtop{\parskip=0pt\parindent=0pt\strut #1\strut}
\eject \nopagenumbers\headline={\hss\tenrm--\ \folio\ --\hss}}
\def\begincaption{\medskip\openup-1\jot\eightpoint}
\def\endcaption{\tenpoint\openup1\jot\leftskip=0pt\rightskip=0pt}
\def\caption#1#2{\message{#1}\begincaption\leftskip=15true mm\rightskip=15true
mm\vbox{\halign{\vtop{\parindent=0pt\parskip=0pt\strut##\strut}\cr{\bf#1}\quad
#2\cr}}\endcaption}

\font\sixrm=cmr6 \font\sixi=cmmi6 \font\sixsy=cmsy6 \font\sixbf=cmbx6
\font\eightrm=cmr8 \font\eighti=cmmi8 \font\eightsy=cmsy8 \font\eightbf=cmbx8
\font\eighttt=cmtt8 \font\eightit=cmti8 \font\eightsl=cmsl8 \font\ninerm=cmr9
\font\ninesy=cmsy9 \font\bold=cmbx10 scaled\magstep1 
\font\sevenit=cmti7
\def\tenpoint{\def\rm{\fam0\tenrm} \textfont0=\tenrm \scriptfont0=\sevenrm
\scriptscriptfont0=\fiverm \textfont1=\teni \scriptfont1=\seveni
\scriptscriptfont1=\fivei \textfont2=\tensy \scriptfont2=\sevensy
\scriptscriptfont2=\fivesy \textfont3=\tenex \scriptfont3=\tenex
\scriptscriptfont3=\tenex \textfont\itfam=\tenit \def\it{\fam\itfam\tenit}
\textfont\slfam=\tensl \def\sl{\fam\slfam\tensl} \textfont\ttfam=\tentt
\def\tt{\fam\ttfam\tentt} \textfont\bffam=\tenbf \scriptfont\bffam=\sevenbf
\scriptscriptfont\bffam=\fivebf \def\bf{\fam\bffam\tenbf}
\setbox\strutbox=\hbox{\vrule height8.5pt depth3.5pt width0pt}
\let\sc=\eightrm \let\big=\tenbig \rm} \def\eightpoint{\def\rm{\fam0\eightrm}
\textfont0=\eightrm \scriptfont0=\sixrm \scriptscriptfont0=\fiverm
\textfont1=\eighti \scriptfont1=\sixi \scriptscriptfont1=\fivei
\textfont2=\eightsy \scriptfont2=\sixsy \scriptscriptfont2=\fivesy
\textfont3=\tenex \scriptfont3=\tenex \scriptscriptfont3=\tenex
\textfont\itfam=\eightit \def\it{\fam\itfam\eightit} \textfont\slfam=\eightsl
\def\sl{\fam\slfam\eightsl} \textfont\ttfam=\eighttt
\def\tt{\fam\ttfam\eighttt} \textfont\bffam=\eightbf \scriptfont\bffam=\sixbf
\scriptscriptfont\bffam=\fivebf \def\bf{\fam\bffam\eightbf}
\setbox\strutbox=\hbox{\vrule height7pt depth2pt width0pt} \let\sc=\sixrm
\let\big=\eightbig \rm} \def\sevenpoint{\def\rm{\fam0\sevenrm}
\textfont0=\sevenrm \scriptfont0=\fiverm \scriptscriptfont0=\fiverm
\textfont1=\seveni \scriptfont1=\fivei \scriptscriptfont1=\fivei
\textfont2=\sevensy \scriptfont2=\fivesy \scriptscriptfont2=\fivesy
\textfont3=\tenex \scriptfont3=\tenex \scriptscriptfont3=\tenex
\textfont\itfam=\sevenit \def\it{\fam\itfam\sevenit} \textfont\bffam=\sevenbf
\scriptfont\bffam=\fivebf \scriptscriptfont\bffam=\fivebf
\def\bf{\fam\bffam\sevenbf} \setbox\strutbox=\hbox{\vrule height6.5pt depth1.5
pt width0pt} \let\sc=\fiverm \let\big=\sevenbig \rm} \def\tenbig#1{{\hbox
{$\left#1\vbox to8.5pt{}\right.\n@space$}}} \def\eightbig#1{{\hbox{$\textfont0
=\ninerm \textfont2=\ninesy \left#1\vbox to 6.5pt{}\right.\n@space$}}}
\def\sevenbig#1{{\hbox{$\textfont0=\eightrm \textfont2=\eightsy\left#1\vbox to
5.5pt{}\right.\n@space$}}}


\def\rarr{\rightarrow}\def\darr{\leftrightarrow} \def\br{\hfil\break}
\mathchardef\hash="015D
\def\U#1{$\underline{\hbox{#1}}$} \def\dsp{\displaystyle}\def\scr{\scriptstyle}
\def\sqr#1#2#3{{\vbox{\hrule height.#2pt \hbox{\vrule width.#2pt
height#1pt\kern#1pt\vrule width.#2pt}\hrule height.#2pt}\hbox{\hskip.#3em}}}
\def\Dal{\,{\mathchoice\sqr64{15}\sqr64{15}\sqr431\sqr331}} \def\et{\eta}
\def\goesas{\mathop{\sim}\limits} \def\Y#1{^{\raise2pt\hbox{$\scr#1$}}}
\def\Z#1{_{\lower2pt\hbox{$\scr#1$}}} \def\const{\hbox{const.}}
\def\al{\alpha}\def\be{\beta}\def\ga{\gamma}\def\de{\delta}\def\ep{\epsilon}
\def\ee{\varepsilon}\def\ka{\kappa}\def\th{\theta}\def\ph{\phi}
\def\ch{\chi}\def\varch{{\raise.4516ex\hbox{$\chi$}}}\def\la{\lambda}
\def\rh{\rho}\def\si{\sigma}\def\SI{\Sigma}\def\pt{\partial}
\def\DE{\Delta}\def\ta{{\tau}}\def\OO{{\rm O}}\def\A{{\cal A}}
\def\OM{\Omega}\def\dd{{\rm d}}
\def\LA{\Lambda}\def\ze{\zeta} \def\GA{\Gamma}
\def\PL#1{Phys.\ Lett.\ {\bf#1}} \def\CMP#1{Commun.\ Math.\ Phys.\ {\bf#1}}
\def\PRL#1{Phys.\ Rev.\ Lett.\ {\bf#1}} \def\MPL#1{Mod.\ Phys.\ Lett.\ {\bf#1}}
\def\AP#1#2{Ann.\ Phys.\ (#1) {\bf#2}} \def\PR#1{Phys.\ Rev.\ {\bf#1}}
\def\CQG#1{Class.\ Quantum Grav.\ {\bf#1}} \def\NP#1{Nucl.\ Phys.\ {\bf#1}}
\def\GRG#1{Gen.\ Relativ.\ Grav.\ {\bf#1}} \def\NC#1{Nuovo Cimento {\bf#1}}
\def\JP#1{J.\ Phys.\ {\bf#1}} \def\AIHP#1{Ann.\ Inst.\ H. Poincar\'e {\bf#1}}
\def\JMP#1{J.\ Math.\ Phys.\ {\bf#1}}
\def\Do{Department of\ }\def\DoP{\Do Physics}
\def\Un{University}\def\Uno{\Un\ of\ }

\def\KK{Kaluza-Klein}\def\ddim{$D$-dimensional\ } \def\g#1{{\rm g}\Z#1}
\def\gh{\hat g}\def\gb{\bar g}\def\Rh{\hat R}\def\V{{\cal V}}\def\rH{\hat r}
\def\ad#1{a\Z{#1}} \def\ld#1{\la\Z{#1}} \def\tee{\tilde\ee} \def\HDE{\hat\DE}
\def\sig{\exp\left(2\ka\si\over\sqrt{D-1}\right)} \def\sqm{\sqrt{m+1}}
\def\sigg{\left(1-\ee\sig\right)} \def\ff{\left[4\ee\ka^2f'(R)\right]}
\def\sigo{\exp\left(-2D\ka\si\over(D-2)\sqrt{D-1}\right)} \def\xb{\bar x}
\def\gxxh{\hat g_{ab}\dd\hat x^a\dd\hat x^b}\def\uu{{\hat u}}\def\vv{{\hat v}}
\def\eu{e^\uu} \def\ev{e^\vv} \def\euu{e^{2\uu}} \def\evv{e^{2\vv}}
\def\CT{\tilde C} \def\CH{\hat C} \def\mG#1{\g#1^{\ 2}-m-1} \def\lb{\bar\la}
\def\gxxb{\gb_{\al\be}\dd\xb^\al\dd\xb^\be} \def\lbz{\lb e^{2\ze}}
\def\laet{\ld1e^{2\et}} \def\lach{\ld2e^{2\varch}} \def\lab{\la_ie^{2(\al_i
\varch+\be_i\et)}} \def\ssum{\sum_{i=3}^s} \def\Mg#1{m+1-\g#1^{\ 2}}
\def\MgB#1{\left(\Mg#1\right)} \def\Mgg#1#2{m+1-\g#1\g#2} \def\GG{\g1^{\ 2}}
\def\MggB#1#2{\left(\Mgg#1#2\right)} \def\rhb{\bar\rh} \def\cd#1{c\Z{#1}}
\def\mg#1{1+(m-1)\g#1^{\ 2}} \def\mgB#1{\left(\mg#1\right)} \def\gd{\g 1-\g 2}
\def\mgg#1#2{1+(m-1)\g#1\g#2} \def\mggB#1#2{\left(\mgg#1#2\right)}
\def\gg#1#2{\left(\g#1\g#2-1\right)} \def\gs#1{\left(\g#1^{\ 2}-1\right)}
\def\lz1{\ld1Z^2} \def\lw2{\ld2W^2} \def\lwzi{\la_iW^{2\al_i}Z^{2\be_i}}
\def\lwzt{\ld3W^{4/3}Z^{2/3}} \def\lwzr{(\gd)\ld4Z^{4/3}\GA^{2/3}}
\def\st{\sin\th} \def\sp{\sin\ph} \def\ct{\cos\th} \def\cp{\cos\ph}
\def\mC{{1\over2}(m-1)C} \def\expm{\exp\left[\mC\xi\right]} \def\bd#1{b\Z{#1}}
\def\Dd#1{\DE_{#1}} \def\CB{\bar C} \def\kb{\bar k} \def\Cb{{1\over2}\CB}
\def\expc{\exp\left[\Cb\xi\right]} \def\DEB{\bar\DE} \def\LEB{\bar\LA\Z1}
\def\HCBM{{1\over2}|\CB|} \def\cosxi{\cos\left(\CT(\xi-\xi\Z0)\right)}
\def\lccos{\LEB^{\ 1/2}|\CT|^{-1}\cosxi} \def\ccurve{$\lb=W=Z=0$ curve}
\def\GGG{\g2^{\ 2}} \def\A{{\cal A}} \def\sap{\ad ppR^{p-1}} \def\Rca{R_{;a}}
\def\spa{\sum_{p=2}^k} \def\gdx{\sqrt{-g}\,\dd^4 x} \def\psa{\ad pp(R^{p-1})}
\def\lkm{(l_{[a;b}k_c m_{d]})^{;d}} \def\lag{\LA\Z g} \def\kab{k_b m_{c]}}
\def\OZ{O(Z_0)} \def\Gag{\left[{\ld4\over\ld2}\pm\ga^{4/3}\right]}
\def\teega{\left(\tee\ga^{2/k}\right)} \def\elp{\ell\Z{\hbox{\sevenrm Planck}}}
\def\doc{domain of outer communications} \def\rsq{$R+aR^2$ theory}
\def\rsqfd{\rsq\ in four dimensions} \hyphenation{Schwarz-schild}
\def\bech{\left[{b\ee\over|b\ee|}+{\ld4\over\ld2}e^{2(\et-\varch)}\right]}
\def\lacch{\lach\bech^{3/2}} \def\lcross{\ld3e^{2(\et+2\varch)/3}}
\def\lccross{(\gd)\ld4e^{2(2\et+\varch)/3}\bech^{1/2}} \def\Ssc{{\cal M}}
\def\gec{\exp\left[2(\g1\ch-\g2\et)\over\gd\right]}
\def\dVds{{\dd\V\over\dd\si}} \def\dta#1{{\dd#1\over\dd\ta}}


\title{Black holes in higher derivative gravity theories}\vskip-5true mm
\preprintno{ADP-91-168/T104\cr NCL-91 TP9} \footdate{December, 1991; Revised
April, 1992} \authors{Salvatore Mignemi$^{\dag\S\spadesuit}$ and David L.
Wiltshire$^{\ddag\clubsuit}$}
\address{$\dag$ \DoP, \Uno Newcastle-Upon-Tyne,\cr Newcastle-Upon-Tyne NE1 7RU,
United Kingdom.\cr \rm{and}\cr INFN, Sezione di Cagliari, Via Ada Negri 18,
09127 Cagliari, Italy.}\vskip-5true mm \address{$\ddag$ \DoP\ and Mathematical
Physics, \Uno Adelaide,\cr GPO Box 498, Adelaide, S.A. 5001, Australia.}
\vskip-12true mm\openup-2pt\abstract\vskip-5true mm

We study static spherically symmetric solutions of Einstein gravity plus an
action polynomial in the Ricci scalar, $R$, of arbitrary degree, $n$, in
arbitrary dimension, $D$. The global properties of all such solutions are
derived by studying the phase space of field equations in the equivalent theory
of gravity coupled to a scalar field, which is obtained by a field redefinition
and conformal transformation. The following uniqueness theorem is obtained:
provided that the coefficient $\ad2$ of the $R^2$ term in the Lagrangian
polynomial is positive then the only static spherically symmetric
asymptotically flat solution with a regular horizon in these models is the
Schwarzschild solution. Other branches of solutions with regular horizons,
which are asymptotically anti-de Sitter, or de Sitter, are also found. An exact
Schwarzschild-de Sitter type solution is found to exist in the \rsq\ if $D>4$.
If terms of cubic or higher order in $R$ are included in the action, then such
solutions also exist in four dimensions. The general Schwarzschild-de Sitter
type solution for arbitrary $D$ and $n$ is given. The fact that the
Schwarzschild solution in these models does not coincide with the exterior
solution of physical bodies such as stars has important physical implications
which we discuss. As a byproduct, we classify all static spherically symmetric
solutions of \ddim gravity coupled to a scalar field with a potential
consisting of a finite sum of exponential terms.\openup2pt
\endpagefoot{$\S$ Present address: Laboratoire de Physique
Th\'eorique, Institut Henri Poincar\'e, 11, rue P. et M. Curie, 75231 Paris
Cedex 05, France.\br $\spadesuit$ Email: mignemi@ccr.jussieu.fr\hfill
$\clubsuit$ Email: dlw@physics.adelaide.edu.au} \section 1. Introduction

Theories of gravity involving higher powers of the Riemann tensor in the
lagrangian have been proposed in several different contexts since the first
days of general relativity. They were first introduced by Weyl in his affine
theory, which aimed to unify gravity and electromagnetism [1]. Such models
have become attractive again in recent years following the demonstration that
the addition to the Einstein-Hilbert lagrangian of terms quadratic in the Ricci
tensor leads to a renormalisable theory [2]. Unfortunately a massive spin 2
``ghost'' is present in the linearised spectrum, leading to an instability of
the theory and a loss of unitarity. It has been suggested that this
problem would disappear in a full non-perturbative treatment of the model, or
even that the ghost states in the perturbative expansion could be a gauge
artifact [3]. However, a non-perturbative formulation is still a distant goal,
while the latter possibility seems to have been ruled out [4].

The effects of higher derivative gravity have also proven to be useful in
cosmology, beginning with the early work of Starobinsky [5] and Kerner
[6] who introduced higher derivative terms with a view to obtaining solutions
which avoid the initial singularity. Later on it was realised that such models
can lead to inflationary expansion driven only by gravity [7-9]. Higher
derivative models have also been studied in the context of quantum cosmology
[10]. In particular, it has been argued that the introduction of quadratic
terms may solve some of the problems due to the non-positiveness of the
ordinary Einstein-Hilbert action for euclidean quantum cosmology [11]. It is
also interesting to note that quadratic corrections to the action are obtained
from quantum wormhole effects [12].

Finally, we should mention that higher order lagrangians arise naturally in
higher dimensional theories, such as Kaluza-Klein and string models. In the
first case they are introduced in order to obtain spontaneous compactification
from purely gravitational higher-dimensional theories, and
in this context the ghost-free Gauss-Bonnet actions have attracted much
interest [13,14]. In the second case, they are obtained as an effective
low energy action [15].

While the cosmology of these models has been largely studied, both in four
[5-9,16] and higher dimensions [17,18], comparatively little is known
about black hole solutions. The weak field limit has been studied to some
extent for the \rsqfd\ [19,20], but the properties of the full solutions in
higher derivative theories remain largely unexplored. On dimensional grounds,
one would expect that the higher derivative terms become dominant in the
proximity of the singularity. However, it is still possible that horizons exist
and that, contrary to the assumptions of refs.\ [19] and [20], analogues of the
usual uniqueness theorems for stationary axisymmetric black holes can be
derived. Indeed, such conclusions can be immediately drawn in the case of the
\rsqfd\ as a result of the ``no hair'' theorem proved by Whitt [21], as
we show in Appendix A. Thus the question of the nature of the static
spherically symmetric solutions of more general higher derivative models is
very interesting from the point of view of general relativity.

One of the reasons for the lack of attention to the black hole problem is
undoubtedly the difficulty of solving the higher order differential equations
arising in higher derivative models. Some progress can be made, however, by
using the fact that higher derivative theories are equivalent, by redefinition
of the metric, to ordinary Einstein gravity coupled to a scalar plus a massive
spin-2 field. This result was first proven by Higgs [22], and later
rediscovered by Whitt [21], in the case of the \rsqfd, for which only the
scalar field is present in the effective theory.
More recently the equivalence has been extended to the case of actions
containing powers of the Ricci and Riemann tensor [23]\foot{$^{\hash1}
$}{For a different approach to the problem see [24].}. The importance of this
equivalence is that one can now use the formalism of ordinary general
relativity to study the more general higher derivative theories.

Our analysis in this paper will make use of the equivalence of the general
$D$- dimensional action
$$S=\int\dd^{D}x{\sqrt{-g}f(R)\over4\ka^2}\,\eqno(1.1)$$
to Einstein gravity coupled to a scalar field [23]. Here $f$ is an
arbitrary function of the Ricci scalar $R$, and $\ka^2$ denotes the
gravitational constant in $D$ dimensions.  If we define $\si$ by
$${2\ka\si\over\sqrt{D-1}}=\ln\ff,\eqno(1.2a)$$
where
$$\ee=\cases{1,&if $f'>0$,\cr -1,&if $f'<0$,\cr}\eqno(1.2b)$$
and make a conformal transformation
$$\gh_{ab}=\ff^{2/(D-2)}g_{ab},\eqno(1.3)$$
then the field equations derived from (1.1) are equivalent to those derived
from the action
$$\hat S=\int\dd^D\hat x\sqrt{-\gh}\left\{ {\Rh\over4\ka^2}-{1\over D-2}\gh^{
ab}\pt_a\si\pt_b\si-\V(\si)\right\},\eqno(1.4a)$$
where
$$\V={\ee\over4\ka^2}\ff^{-D/(D-2)}\left[Rf'-f\right],\eqno(1.4b)$$
with $R$, $f$ and $f'$ defined implicitly in terms of $\si$ via (1.2).
For the quadratic theory, for example, with $f=R+aR^2$ we find [18]
$$\V={\ee\over16\ka^2a}\exp\left(2(D-4)\ka\si\over(D-2)\sqrt{D-1}\right)\left[
1-\ee\exp\left(-2\ka\si\over\sqrt{D-1}\right)\right]^2,\eqno(1.5)$$
Similarly, for the cubic theory with $f=R+aR^2+bR^3$ we find [16]
$$\eqalign{\V={\ee\over54\ka^2b^2}\sigo\Biggl\{\pm&\left[a^2-3b\sigg\right]^
{3/2}\cr&\qquad-a^3+{9\over2}ab\sigg\Biggr\},\cr}\eqno(1.6)$$
For other actions polynomial in $R$
$$f(R)=R+\spa\ad pR^p,\eqno(1.7)$$
and $f'$ can be inverted to give a precise analytic expression for $R$ in terms
of $\si$, and hence for $\V(\si)$, for a general polynomial only if $n\le5$.

The general higher order field equations obtained from the action (1.1), with
$f$ of the form (1.7), are given by
$$\eqalign{R_{ab}-{1\over2}g_{ab}R+\spa\ad p\Biggl\{&pR^{p-1}R
_{ab}-p(p-1)R^{p-3}\Bigl[RR_{;ab}+(p-2)R_{;a}R_{;b}\Bigr]\cr&+g_{
ab}\left[p(p-1)R^{p-3}\Bigl(R\Dal R+(p-2)R^{;c}R_{;c}\Bigr)-{1\over2}
R^p\right]\Biggr\}=0.\cr}\eqno(1.8)$$
Since the arbitrary-dimensional Schwarzschild solution
$$\dd s^2=-\left(1-{2GM\over r^{D-3}}\right)\dd t^2+\left(1-{2GM\over r^{D-3}}
\right)^{-1}\dd r^2+\dd\OM^2_{D-2},\eqno(1.9)$$
has $R=0$ throughout the domain of outer communications, it is clear that the
Schwarz\-schild solution solves the equations (1.8) for any choice of the
constants $\ad p$. Thus the problem before us is to determine whether the
Schwarzschild solution is the only static spherically symmetric asymptotically
flat solution with a regular horizon. On account of the equivalence of the
higher order theory to the theory described by the action (1.4) this uniqueness
problem can be regarded to as the problem of establishing a ``no hair theorem''
for the latter model.

Whitt established such a theorem in the case of the \rsqfd\foot{$^{\hash2}$}{In
fact, a black hole uniqueness theorem for the more restricted case of pure
$R^2$ theory, without an Einstein-Hilbert term, was obtained much earlier on by
Buchdahl [25].} [21], by demonstrating that all asymptotically flat
stationary axisymmetric solutions to the higher order vacuum  equations must
have $R=0$ in the \doc. If one adds extra matter fields to the action one still
finds that $R=0$ for stationary, axisymmetric, asymptotically flat solutions
provided that the energy-momentum tensor is traceless and satisfies the matter
circularity condition. For such solutions, therefore, (1.8) becomes equivalent
to the usual Einstein equations, and the usual uniqueness theorems and
no hair theorems will carry over to the fourth order theory. If one considers
an arbitrary polynomial in $R$ of the form (1.7), however, then Whitt's
argument breaks down, as we demonstrate in Appendix A. Consequently a
different approach is called for.

Our approach here will differ not only from that of Whitt, but also from other
standard approaches to no hair theorems [26,27], in that we will solve the
problem by studying the phase space of the field equations obtained from
(1.4). We will take advantage of the fact that by making a judicious choice of
coordinates these equations may be written in the form of a 5-dimensional
autonomous system of ordinary first-order differential equations, so that all
the global properties of the solutions can be derived. Our approach not only
has the advantage that it can be used to establish a black hole uniqueness
theorem for general actions polynomial in $R$, but it will also enable us to
determine the nature of other solutions in these models which have regular
horizons but which are not asymptotically flat.

The analysis we will use here is very similar to that developed in refs.\ [28]
and [29], henceforth denoted I and II respectively, where we derived the global
properties of static spherically symmetric solutions in models of gravity which
arise from the dimensional reduction of certain higher-dimensional gravity
theories. The scalar field in (1.4) then corresponds to the radius of the extra
dimensions, (the ``compacton''), and the potential $\V(\si)$ contains one or
two exponential terms.

The analysis of I and II is based on the fact that the appropriate field
equations can be reduced to a 5-dimensional autonomous system of first-order
ordinary differential equations. This is possible essentially due to the fact
that the field equations form a system very similar to those of a Toda lattice
when written in terms of
appropriate coordinates [30], (the Toda lattice being an integrable system).
Since the metric and scalar fields are related to the functions $X$, $Y$, $V$,
$Z$ and $W$ of the 5-dimensional phase space, $\Ssc$, they are necessarily
regular at all points of the integrals curves apart from critical points.
Consequently, in order to determine the global properties of all solutions --
namely the structure of their singularities, horizons and asymptotic regions --
it suffices to study the properties of the solutions at critical points of
$\Ssc$. In order to determine which critical points are connected to which
other ones by integral curves requires a careful analysis of the structure of
the space $\Ssc$: in particular of surfaces corresponding to particular
subspaces, which separate integral curves corresponding to spacetimes of
different causal structures. This method represents a powerful analytical tool
for these (and possibly other) models for which it is not possible to write
down the general static spherically symmetric solutions in a closed analytic
form.

One should mention that on account of the way in which the phase space
functions are constructed, the metric functions (with signature $-++\dots+$)
are necessarily positive. Thus for the Schwarzschild solution the integral
curves obtained correspond to the \doc\ only. A similar analysis of the
Reissner-Nordstr\"om solution yields a phase space with distinct regions which
correspond to (i) integral curves in the \doc; and (ii) integral curves in
the region between the singularity and the inner horizon. (These two regions
of the phase space are separated by a surface which corresponds to the
Robinson-Bertotti solutions.) Thus the method described does not pick out
regions of the spacetimes in which the Killing vector $\pt/\pt t$ is
spacelike. Such regions can only be obtained by continuation of such solutions
as are described here.

The first step of our analysis here will be to generalise the work of I and II
to establish the properties of solutions to the equations derived from (1.4)
with a potential consisting of a sum of $s$ exponential terms
$$\V(\si)={-1\over4\ka^2}\;\sum_{i=1}^s\la_i\exp\left(-4\g i\ka\si\over D-2
\right),\eqno(1.10)$$
the $\la_i$ and $\g i$, $i=1,\dots,s$ being constants. Of the higher derivative
theories, only the \rsq\ has an action which is precisely of this
form. However, for each $n$ there will be a special choice of the constants
$\ad p$ for which the potential (1.4b) reduces to this form. This will be so if
the $(n-1)$-th root of $f'$ is linear in $R$. In particular, if we choose
$$\ad p={(2\ad2)^{p-1}(n-2)!\over(n-1)^{p-2}p!(n-p)!}\,,\qquad 3\le p\le n,
\eqno(1.11)$$
then
$$f'=\left(1+{2\ad2R\over n-1}\right)^{n-1},\eqno(1.12)$$
and we find that the potential takes the form
$$\eqalign{\V={(n-1)^2\over8n\ka^2\ad2}\sigo\Biggl\{&\tee\exp\left(2n\ka\si
\over(n-1)\sqrt{D-1}\right)\cr&\qquad-{n\over n-1}\sig+{\tee^{n-1}\over n-1}
\Biggr\},\cr}\eqno(1.13a)$$
where\foot{$^{\hash3}$}{Note that $\ee=\tee^{n-1}$ with these definitions: for
$n$ even $\ee$ and $\tee$ have the same sign. However, if $n$ is odd then $\ee
=+1$, and the sign of $\tee$ corresponds instead, for example, to the plus or
minus sign multiplying the first term in (1.6).}
$$\tee=\cases{1,&if $1+2\ad2R/(n-1)>0$,\cr -1,&if $1+2\ad2R/(n-1)<0$.\cr}
\eqno(1.13b)$$
We shall call this class of models ``special polynomial $R$ theories''.

Before proceeding further, we note that the definition of asymptotic flatness
in our analysis requires some clarification. On account of the conformal
transformation (1.3) the radial coordinate of the higher derivative theory,
$r$, is related to the radial coordinate of the effective theory, $\rH$ by
$$r=\ff^{-1/(D-2)}\rH=\exp\left(-2\ka\si\over(D-2)\sqrt{D-1}\right)\rH.
\eqno(1.14)$$
Thus solutions which are asymptotically flat in the higher derivative theory
need not be asymptotically flat in the effective theory, since it is the
dependence on $r$ as measured by the metric $g_{ab}$ which is of physical
importance rather than the dependence on $\rH$ as measured by the metric $\gh_
{ab}$. Moreover, the definition of the asymptotic region can vary between
the physical and the effective theories. The definitions of spatial infinity
and of asymptotic flatness will only coincide when $\si\rarr\const$ as $\rH
\rarr\infty$. However, since $\si$ is not a physical field here, (unlike the
Kaluza-Klein case [28,29]), we need place no requirements on its
asymptotic form at spatial infinity in the effective theory. Thus a
discussion of the existence of static spherically symmetric black holes in the
higher derivative models requires an examination of {\it all} static
spherically symmetric solutions in the effective theory, and not merely the
ones which are asymptotically flat according to the effective theory.

We will begin by deriving the global properties for the theory with the
potential (1.10) -- almost all cases are dealt with in \S2. For special values
of the parameters, which include the cases of the \rsq\ and all
potentials (1.13), the structure of the phase space is modified. Such solutions
will be discussed in \S3. We will then generalise the analysis to include the
potential (1.6) to that of the $R+aR^2+bR^3$ theory in \S4. This generalisation
will be found to produce only minor changes to the results of \S2 and \S3. We
will further demonstrate that our results can be extended to the case of the
polynomial of arbitrary degree. Some physical implications of our work are
discussed in \S5.
\section 2. The general exponential sum potential

\subsection 2.1 The dynamical system

As in I and II we will begin by choosing coordinates
$$\gxxh=\euu\left(-\dd t^2+\rH^{2m}\dd\xi^2\right)+\rH^2\gxxb,\eqno(2.1a)$$
where $m=D-2$, $\uu=\uu(\xi)$, $\rH=\rH(\xi)$ and $\gb_{ab}$ is the metric on
an arbitrary $m$-dimensional Einstein space:
$$\bar R_{\al\be}=(m-1)\lb\gb_{\al\be}.\eqno(2.1b)$$
Of course we are most interested in the case in which $\gb_{ab}$ is the metric
on a 2-sphere (and $\lb=1$). However, the structure of the phase space is
better revealed by taking the more general ansatz (2.1b). In order to write the
equations as a first order system which is everywhere well-defined, we will
choose an ordering of the $\g i$ such that
$$\g 1>\g 3>\g 4>\dots>\g s>\g 2.\eqno(2.2)$$
If we now define the functions\foot{$^{\hash4}$}{These functions correspond, in
fact, to the differences of the Toda lattice coordinates [30].} $\ze$, $\et$
and $\ch$ by
$$\ze=\uu+(m-1)\ln\rH\,,\eqno(2.3)$$
$$\et=\uu+m\ln\rH-{2\g 1\ka\si\over m}\,,\eqno(2.4)$$
$$\ch=\uu+m\ln\rH-{2\g 2\ka\si\over m}\,.\eqno(2.5)$$
then the field equations become
$$\ze''=(m-1)^2\lbz+\laet+\lach+\ssum\lab,\eqno(2.6a)$$
$$\eqalign{\et''=m(m-1)\lbz&+{1\over m}\MgB1\laet+{1\over m}\MggB12\lach\cr&+
{1\over m}\ssum\MggB1i\lab,\cr}\eqno(2.6b)$$
$$\eqalign{\ch''=m(m-1)\lbz&+{1\over m}\MggB12\laet+{1\over m}\MgB2\lach\cr&+
{1\over m}\ssum\MggB2i\lab,\cr}\eqno(2.6c)$$
with the constraint
$$\eqalign{(m+1)\ze'^2+{2m\ze'(\g 2\et'-\g 1\ch')\over\gd}+{\mgB2\over(\gd)^2}
{\et'}^2-2{\mggB12\over(\gd)^2}\et'\ch'&\cr+{\mgB1\over(\gd)^2}{\ch'}^2+(m-1)
\lbz+{\ld1\over m}e^{2\et}+{\ld2\over m}e^{2\varch}+{1\over m}\ssum\lab&=0,\cr}
\eqno(2.6d)$$
where, for $s\ge3$,
$$\al_i={\g 1-\g i\over\gd},\qquad i=3,\dots,s,\eqno(2.6e)$$
and
$$\be_i=1-\al_i={\g i-\g 2\over\gd},\qquad i=3,\dots,s,\eqno(2.6f)$$
If $s<3$ then the last summation term in (2.6d) vanishes. Note that on account
of (2.2), $0<\al_i<1$, $0<\be_i<1$, and $\al_i=\be_i$ only in the special case
in which $s=3$ and $\g 3=\left(\g 1+\g 2\right)/2$.

These equations can be recast in the form of a 5-dimensional autonomous
system of first-order differential equations. If we define variables $V$, $W$,
$X$, $Y$ and $Z$ by
$$\eqalign{&V=\ch',\qquad W=e^{\ch},\qquad X=\ze',\qquad Y=\et',\qquad
Z=e^{\et},\cr}\eqno(2.7)$$
then the constraint (2.6d) can be regarded as a definition of $e^{2\ze}$.
Eliminating the $e^{2\ze}$ terms from (2.6a-c) we therefore obtain the system
\foot{$^{\hash5}$}{We are using a different normalisation here for $Z$ and $W$
compared with that used in I and II.}
$$X'={1\over m}\left\{\lz1+\lw2+\ssum\lwzi\right\}-{(m-1)P\over m},\eqno(2.8a)
$$
$$Y'={-1\over m}\left\{\gs1\lz1+\gg12\lw2+\ssum\gg1i\lwzi\right\}-P,
\eqno(2.8b)$$
$$V'={-1\over m}\left\{\gg12\lz1+\gs2\lw2+\ssum\gg2i\lwzi\right\}-P,
\eqno(2.8c)$$
$$Z'=YZ,\eqno(2.8d)$$
$$W'=VW,\eqno(2.8e)$$
where
$$\eqalign{P\equiv m\Biggl\{(m+1)X^2+&{2mX(\g 2Y-\g 1V)\over\gd}+{\mgB2Y^2\over
(\gd)^2}\cr&\qquad-{2\mggB12YV\over(\gd)^2}+{\mgB1V^2\over(\gd)^2}\Biggr\}.\cr}
\eqno(2.8f)$$

The dynamical system of the Kaluza-Klein models dealt with in II, with $n_e$
extra dimensions and a higher-dimensional cosmological constant $\LA$, is
retrieved by setting $s=2$ and
$$\g1=\g2^{\ -1}={\sqrt{m+n_e}\over\sqrt{n_e}},\qquad\ld1=n_e(n_e-1)\tilde\la,
\qquad\ld2=-2\LA.\eqno(2.9)$$

In the case of the \rsq, and indeed any higher order theory given by
the potential (1.13), the appropriate dynamical system is obtained by taking
equations (2.8) and setting $s=3$,
$$\g1={m+2\over2\sqm},\qquad\g2={2(n-1)-m\over2(n-1)\sqm},\qquad\g3={1\over\sqm
},\eqno(2.10a)$$
$$\al\Z3={n-1\over n},\qquad\be\Z3={1\over n},\eqno(2.10b)$$
and
$$\ld1={-\tee^{n-1}(n-1)\over2n\ad2},\qquad\ld2={-\tee(n-1)^2\over2n\ad2},
\qquad\ld3={(n-1)\over2\ad2}\,.\eqno(2.10c)$$

As in the case of the \KK\ models the phase space has a great many symmetries
which greatly simplify the analysis. Equations (2.8b) and (2.8e) ensure that
trajectories cannot cross either the $W=0$ or the $Z=0$ subspaces. These two
subspaces correspond physically to the cases in which $\ld2=0$ and $\ld1=0$
respectively. As we have written them, equations (2.8) are valid for $W\ge0$
and $Z\ge0$. It is possible to make the equations valid for all $W$ and for all
$Z$ by introducing modulus signs in the terms $W^{2\al_i}$ and $Z^{2\be_i}$
which involve fractional powers of $Z$ and $W$. However, this merely introduces
a trivial symmetry between trajectories in the $W>0$ and $W<0$ portions of the
phase space, and between trajectories in the $Z<0$ and $Z>0$ portions of the
phase space. Thus we may restrict our attention to $Z\ge0$ and $W\ge0$ without
loss of generality.

The hyperboloid defined $\lb=0$, or
$$P+\ld1Z^2+\ld2W^2+\ssum\lwzi=0,\eqno(2.11)$$
similarly forms a surface which trajectories cannot cross. It partitions the
phase space into the two physically distinct regions with $\lb>0$ and $\lb<0$.

If $W=0$ and $\g1\ne0$ then
$$V={\g1(\gd)(mX+\cd1)+\mggB12 Y\over\mg1},\eqno(2.12)$$
while if $Z=0$ and $\g2\ne0$ then
$$Y={-\g2(\gd)(mX+\cd2)+\mggB12 V\over\mg2},\eqno(2.13)$$
where $\cd1$ and $\cd2$ are arbitrary constants. (If $\g1=0$ or $\g2=0$ we have
instead $V=Y+\const$) Physically (2.12) is equivalent to $s=1$, (i.e., $\ld i=0
$, $i\ge2$), while (2.13) gives the same system with $\ld1\rarr\ld2$ and $\g1
\rarr\g2$. Thus in each case a further degree of freedom can be integrated out,
giving rise to a 3-dimensional autonomous system. The properties of such
systems were studied in I and II. Further simplifications arise if in addition
one of the constants  $\lb$ or $\ld1$ (or $\ld2$ as appropriate) is zero. In
these cases it is in fact possible to integrate the field equations exactly.
For completeness we list these solutions in Appendix B.
\subsection 2.2 The (anti)-de Sitter subspaces

In addition to the $W=0$, $Z=0$ and $\lb=0$ subspaces, there is at least one
other 3-dimensional subspace which exists for all $s\ge2$, which was not
studied in detail in II. These subspaces may be identified by noting that
solutions for which $\si$ is constant globally form a special class. The actual
value which this constant takes may be determined from the field equations.
In particular, we find that if $V=Y$ and $W=\ga Z$, ($\ga\ge0$), where
$$\ld1\g1+\ld2\g2\ga^2+\ssum\la_i\g i\ga^{2\al_i}=0,\eqno(2.14)$$
then the field equations (2.8) reduce to the 3-dimensional
system
$$X'=-\LA Z^2-{(m-1)P\over m},\eqno(2.15a)$$
$$Y'=-\LA Z^2-P,\eqno(2.15b)$$
$$Z'=YZ,\eqno(2.15c)$$
where
$$\LA={-1\over\; m}\left(\ld1+\ld2\ga^2+\ssum\la_i\ga^{2\al_i}\right),\eqno
(2.15d)$$
and $P$ now simplifies down to
$$P=m(X-Y)\left((m+1)X-(m-1)Y\right).\eqno(2.15e)$$
In the case $s=2$, for example, (2.14) is satisfied by any $\g1$ and $\g2$
provided that
$$\ga=\left(-\g1\ld1\over\g2\ld2\right)^{1/2}\eqno(2.16)$$
is real. For $s>2$, (2.14) may have more than one solution.

If $s=1$ we of course obtain equations (2.15) if $\g1=0$ and $\ld1=-m\LA$.
Consequently, equations (2.15) are formally equivalent to those of the
Schwarzschild-de Sitter solution in $(m+2)$ dimensions, and can be integrated
explicitly in terms of the coordinate $\rH$. We therefore obtain the solution
$$\gxxh=-\HDE\dd t^2+\HDE^{-1}\dd\rH^2+\rH^2\gxxb,\eqno(2.17a)$$
with
$$\HDE=\lb-{C\over\rH^{m-1}}-{\LA\ga^{-2\g1/(\gd)}\rH^2\over m+1}\,,\eqno
(2.17b)$$
and constant scalar field
$$e^{2\ka\si}=\ga^{m/(\gd)},\eqno(2.17c)$$
$C$ being an arbitrary constant. In the case of the higher order theories with
potentials (1.13), a conformal transformation back to the original metric
yields the solution
$$\dd s^2=-\DE\dd t^2+\DE^{-1}\dd r^2+r^2\gxxb,\eqno(2.18a)$$
with
$$\DE=\lb-{2GM\over r^{m-1}}-{\LA\ga^{-2(n-1)/n}\,r^2\over m+1}\,,\eqno(2.18b)
$$
and $M=C/[2G\ga^{2(m-1)(n-1)/(mn)}]$. (We have rescaled $t$ in obtaining
(2.18).) The solutions are asymptotically de Sitter if $\LA>0$ (or anti-de
Sitter if $\LA<0$). In the case of the \rsq, we have
$$\DE=\lb-{2GM\over r^{m-1}}+{mr^2\over(m+1)(m+2)(m-2)a}\,,\eqno(2.19)
$$
and consequently solutions of this type exist only for $m>2$, i.e., for $D>4$.
These solutions are asymptotically anti-de Sitter if $a>0$. For $n>2$
solutions of the type (2.18) also exist for $D=4$. In general, it is possible
to find more than one branch of solutions -- some of which are asymptotically
de Sitter, and some of which are asymptotically anti-de Sitter.

If $M=0$ we retrieve cosmological de Sitter solutions, the existence of which
has been discussed previously by Barrow and Ottewill [8] for an arbitrary $f(R)
$ lagrangian in four dimensions, and by Madsen and Barrow [31] for more
generalised higher derivative lagrangains in arbitrary dimensions.

The solutions (2.17) do not exhaust all possible solutions to equations (2.15).
In particular, a special class of solutions for which $\rH(\xi)$ is everywhere
constant are also admitted. These solutions may be determined by direct
integration of (2.15) in the case that $Y=X$. Since $P=0$ in this instance we
have $Y'=X'$ and (2.15) reduces to a 2-dimensional autonomous system. The
equations can be readily integrated using the coordinate $Z=e^{\et}$, and we
find
$$\gxxh=\ga^{2\g1/(\gd)}\left\{-Z^2\dd t^2+{\dd Z^2\over C-\LA Z^2}+\left((m-1)
\lb\over\LA\right)\gxxb\right\},\eqno(2.20)$$
where $C$ is an arbitrary constant, and $Z>0$. (We have used the freedom of
rescaling $t$ to remove an unphysical constant.) Thus we find a solution with
$\lb>0$ provided that $\LA>0$. These solutions are topologically a product of
2-dimensional anti-de Sitter space with an $m$-sphere of constant curvature
-- a type of Robinson-Bertotti solution. Solutions with $\lb<0$ and $\LA<0$ are
similarly a product of 2-dimensional de Sitter space with $m$-dimensional
hyperbolic space.

To compare these results with our later studies of the 5-dimensional phase
space it is useful also to give a description of the 3-dimensional phase
space here.

Provided that $\LA\ne0$ then the only critical points at a finite distance from
the origin are the lines
$$Z=0,\qquad Y=X,\eqno(2.21)$$
and
$$Z=0,\qquad Y={(m+1)X\over m-1},\eqno(2.22)$$
These solutions have $\lb=0$ as well as $W=Z=0$. Solutions lying in the
$Z=0$ plane, as depicted in Fig.~1, are Schwarzschild solutions. Equations
(2.15) can be integrated directly in this case since
$$Y={m\over m-1}(X+k),\eqno(2.23)$$
where $k$ is an arbitrary constant.
The critical points in the first quadrant are found to correspond to the limit
$\xi\rarr-\infty$, while the critical points in the third quadrant correspond
to the limit $\xi\rarr\infty$. We find that endpoints of trajectories on the
line $Y=X$ correspond to horizons with $\rH\rarr\const$, while endpoints of
trajectories on the line $Y=(m+1)X/(m-1)$ correspond to $\rH\rarr0$
singularities. The $Z=0$ trajectories ending on the line $Y=X$ thus represent
positive mass Schwarzschild solutions, while those ending on the line $Y=(m+1)
X/(m-1)$ represent negative mass Schwarzschild solutions with naked
singularities. The constant $k$ in (2.23) is related to the Schwarzschild mass:
the $k=0$ trajectories represents flat space. Furthermore, if a trajectory
which corresponds to a positive mass Schwarzschild solution in one quadrant is
traced back to the opposite quadrant, then one obtains the negative mass
Schwarzschild solution of the same absolute mass in the $\lb>0$ region.

If $\LA=0$ then additional critical points exist near the origin. Furthermore,
(2.15) reduces to a 2-dimensional autonomous system for all $X$, $Y$ and $Z$.
We will defer discussion of this special subspace until \S3.

Small perturbations about the critical points (2.21) and (2.22) yield the
eigenvalues $\la=0,Y_0,2X_0$ in the 3-dimensional subspace, (where $Y_0=
X_0$ or $Y_0=(m+1)X_0/(m-1)$ as appropriate), and additional eigenvalues
$\la=0,Y_0$ for perturbations in the extra directions in the full
5-dimensional phase space. The pattern of trajectories is therefore
identical to that of the corresponding trajectories in I and II. In the
3-dimensional subspace each critical point $(X_0,Y_0,0)$ in the first
(third) quadrant repels (attracts) a 2-dimensional set of trajectories which
lie approximately in the plane $Y=(mX\mp X_0)/(m-1)$. Since the trajectories
are approximately planar the one zero eigenvalue corresponds to the degenerate
direction perpendicular to this plane. In the 5-dimensional phase space
there is an extra degenerate direction, and the dimension of the set of
trajectories repelled (attracted) for first (third) quadrant trajectories
increases by one.

To complete the description of the phase space it is necessary to describe the
critical points at infinity. Following I and II, we introduce spherical polar
coordinates
$$X=\rh\st\cp\,\qquad Y=\rh\st\sp\,\qquad Z=\rh\ct\,.\eqno(2.24)$$
The surface at infinity is then brought to a finite distance from the origin
by the transformation
$$\rh=\rhb(1-\rhb)^{-1},\qquad0\le\rhb\le1.\eqno(2.25)$$
If we define a coordinate $\ta$ by $\dd\ta=\rh\dd\xi=\rhb(1-\rhb)^{-1}d
\xi$, then on the sphere at infinity, i.e.\ at $\rhb=1$, $\dd\rhb/\dd\ta=0$
identically while
$${\dd\th\over\dd\ta}=\ct\Biggl\{-\LA\cos^2\th\left[\cp+\sp\right]-\sin^2\th
\left[\sp+\bar P\Z1\left({(m-1)\over m}\cp+\sp\right)\right]\Biggr\},\eqno
(2.26a)$$
$${\dd\ph\over\dd\ta}={1\over\st}\Biggl\{-\LA\cos^2\th\left(\cp-\sp
\right)+\sin^2\th\bar P\Z1\left[{(m-1)\over m}\sp-\cp\right]\Biggr\},
\eqno(2.26b)$$
where
$$\bar P\Z1=m\left((m+1)\cos^2\ph-2m\cp\sp+(m-1)\sin^2\ph\right).
\eqno(2.26c)$$
Four sets of critical points are found: \case (i):
First of all we obtain the endpoints of the lines of critical points $Y=X$ and
$Y=(m+1)X/(m-1)$, for which $\lb=W=Z=0$. The points located at
$$\th={\pi\over2},\qquad\ph={\pi\over4},\ {5\pi\over4}\eqno(2.27a)$$
$$\hbox{or}\qquad\qquad X=\pm\infty,\qquad Y=X,\qquad Z=0,\eqno(2.27b)$$
will be denoted $L_1$ and $L_2$. The points located at
$$\th={\pi\over2},\qquad\ph=\arctan\left(m+1\over m-1\right)\eqno(2.28a)$$
$$\hbox{or}\qquad\qquad X=\pm\infty,\qquad Y=\left(m+1\over m-1\right)X,\qquad
Z=0.\eqno(2.28b)$$
will be denoted $L_3$ and $L_4$. As for points at a finite distance from the
origin, the points $L_1$ and $L_2$ correspond to horizons, and $L_3$ and $L_4$
to $\rH\rarr0$ singularities. The points $L_{1,3}$ ($L_{2,4}$) repel (attract)
a 2-dimensional set of trajectories, which are unphysical, however, since
they are confined to the sphere at infinity.
\case (ii): Two critical points, which we will denote $M_1$ and $M_2$, are
located at
$$\th={\pi\over2},\qquad\ph=\arctan\left(m\over m-1\right)\eqno(2.29a)$$
$$\hbox{or}\qquad\qquad X=\pm\infty,\qquad Y=\left(m\over m-1\right)X,\qquad Z
=0,\eqno(2.29b)$$
in the $\lb>0$ portion of the phase space. These points correspond to the
asymptotic region ($\rH\rarr\infty$) of the Schwarzschild solutions which lie
in the $Z=0$ plane. These are the only trajectories which end on these points:
they are found to be saddle points with respect to other directions in the
phase space. \case (iii): If $\LA<0$ then there are two critical points, which
we will denote $S_1$ and $S_2$, which are located at
$$\th=\arctan\sqrt{-2\LA},\qquad\ph={\pi\over4},\ {5\pi\over4}\eqno(2.30a)$$
$$\hbox{or}\qquad\qquad X=\pm\infty,\qquad Y=X,\qquad Z={X\over\sqrt{-\LA}},
\eqno(2.30b)$$
in the $\lb<0$ portion of the phase space. At these points we find that $\rH
\rarr\const$ These points correspond to the $Z\rarr\infty$ region of the
Robinson-Bertotti-like solutions (2.20). An analysis of small perturbations
in the 3-dimensional subspace reveals that the point $S_1$ ($S_2$) in the
first (third) quadrant is an attractor (repellor) for a 2-dimensional set of
trajectories: these are of course the solutions (2.20) with $\LA<0$ and
$\lb<0$, which lie in the plane $Y=X$. This clarifies the nature of the
corresponding points $S_{1-8}$ in II, which were not discussed in detail
there\foot{$^{\hash6}$}{Since we are restricting the analysis to $W\ge0$ and
$Z\ge0$ here, we of course obtain half or quarter as many critical points as
in I and II for those points with $W\ne0$ or $Z\ne0$.}.

If $\LA>0$ (and $\lb>0$) then no  $Z\rarr\infty$ region is defined since the
maximum value $Z$ can take is $(C/\LA)^{1/2}$. Thus instead of ending on the
points $S_{1,2}$ trajectories move from one point on the $Z=0$, $Y=X$ line to
another point on the same line in the opposite quadrant. This pattern is made
clear by Fig.~2 where we plot the $Y=X$ plane through the 3-dimensional
subspace.

\case (iv): If $\LA<0$ then there are two critical points, which we will denote
$T_1$ and $T_2$, which are located at
$$\th=\arctan\left(-(2m^2+2m+1)\LA\over m+1\right)^{1/2},\qquad\ph=
\arctan\left(m+1\over m\right)\eqno(2.31a)$$
$$\hbox{or}\qquad\qquad X=\pm\infty,\qquad Y={(m+1)X\over m},\qquad Z={X\over
m}\left(m+1\over-\LA\right)^{1/2},\eqno(2.31b)$$
in the portion of the phase space with $\lb=0$. Point $T_1$ ($T_2$) in the
first (third) quadrant is found to attract (repel) a 3-dimensional set of
trajectories in the 3-dimensional subspace. These points of course
correspond to the asymptotic ($\rH\rarr\infty$) region of the
Schwarzschild-anti-de Sitter solutions (2.17) if $\lb>0$. $T_{1,2}$ are
endpoints for trajectories both in the $\lb>0$ and $\lb<0$ portions of the
phase space.

If $\LA>0$ then no asymptotic region is defined. Instead the trajectories move
from the $Y=X$ or $Y=(m+1)X/(m-1)$ line in one quadrant to one of these two
lines in the opposite quadrant. Trajectories starting and finishing on the
$Y=X$ line with $\lb>0$ correspond to the region of the positive mass
Schwarzschild-de Sitter solution between the Schwarzschild and de Sitter
horizons. Trajectories with endpoints on both the $Y=X$ and $Y=(m+1)X/(m-1)$
lines represent the region of the negative mass Schwarzschild-de Sitter
solution between $\rH=0$ and the de Sitter horizon. Trajectories cannot have
two endpoints on the $Y=(m+1)X/(m-1)$ line since they cannot cross the $Y=X$
plane.

The pattern of trajectories on the hemisphere at infinity of the 3-dimensional
subspace is sketched in Fig.~3. Although these trajectories are unphysical it
is helpful to sketch them since by continuity arguments they will determine the
behaviour of the physical integral curves which lie within the hemisphere at
infinity but near its surface.
\subsection 2.3 The $W=0$ and $Z=0$ subspaces

These subspaces were discussed in I and II for particular values of $\g1$ and
$\g2$. The properties for general $\g1$ and $\g2$ are similar. If $\LA=0$ the
only critical points at a finite distance from the origin in the full
5-dimensional phase space have both $W=0$ and $Z=0$, and so are common to
both subspaces. The critical points are located at $X=X_0$ and $Y=Y_0$, where
$$|X_0|\ge\left(m-1\over\mg1\right)^{1/2}|\g1\cd1|,\eqno(2.32)$$
and $Y_0$ is given by solving the quadratic equation
$$(m-1)Y_0^{\ 2}-2mX_0Y_0+(m+1-\GG)X_0^{\ 2}+\g1^{\ 2}\cd1^{\ 2}=0,\eqno(2.33)
$$
with $V$ given by (2.12) with $X=X_0$.

Consider the $W=0$ subspace. If $\GG\le m+1$ then the pattern of the
trajectories is the same as in I and II since all critical points lie in the
first and third quadrants. Points with $Y_0>0$ (first quadrant) correspond to
the limit $\xi\rarr-\infty$, and those with $Y_0<0$ (third quadrant) to the
limit $\xi\rarr+\infty$. Each point in the first (third) quadrant repels
(attracts) a 2-dimensional set of trajectories which lie approximately in the
plane $\dsp Y=\left(m\over m-1\right)\left[X\pm\left((1+(m-1)\GG)X_0^{\ 2}-(m-1
)\GG\cd1^{\ 2}\right)^{1/2}\right]$. There is a zero eigenvalue
corresponding to the degenerate direction perpendicular to this plane. In terms
of the coordinate $\rH$ one finds that $\rH\rarr0$ at all critical points
except those for which $\cd1=mk$, which  correspond to horizons. These special
critical points are of course those lying in the $W=0$, $Z=0$, $V=Y$ plane
(c.f.\ \S2.2). Fig.~1 thus represents the plane which bisects the $W=0$, $Z=0$
subspace to pick out the Schwarzschild solutions.

If $\GG>m+1$ then points for which $|X_0|<X_1$, where
$$X_1={|\g1\cd1|\over\sqrt{\mg1}},\eqno(2.34)$$
lie in the first and third quadrants, and have the same properties as for
$\GG<m+1$. If $|X_0|>X_1$, on the other hand, then one critical point lies in
each quadrant. The two points in the first and third quadrants have the same
properties as before. At the critical points in the second and fourth quadrants
we find that $\rH\rarr\infty$. As before there is one degenerate direction
corresponding to a zero eigenvalue. However, both points are now saddle points
with respect to the remaining two directions. Each critical point in the second
(fourth) quadrant repels (attracts) one trajectory from the $\lb>0$ region and
one trajectory from the $\lb<0$ region of the phase space: these are the $Z=0$
solutions discussed in Appendix B.1, and depicted in Fig.~4(b). Each point
in the second (fourth) similarly attracts (repels) one trajectory for each sign
of $\ld1$: these trajectories are in fact the $\lb=0$ solutions discussed in
Appendix B.2, and depicted in Fig.~5(c). Their asymptotic form is given by
(B.27).  Thus trajectories in the $\lb>0$ and $\lb<0$ regions of the phase
space for which $Z$ is not identically zero do not have endpoints in the second
and fourth quadrants. If such solutions have asymptotic regions, the limit $\rH
\rarr\infty$ must be approached at critical points at the phase space infinity.

The phase space infinity of the $W=0$ subspace can be studied
once again by introducing coordinates (2.24), (2.25). The following equations
are obtained for the angular coordinates on the $\rhb=1$ sphere at infinity
$$\eqalign{{\dd\th\over\dd\ta}=\ct\Biggl\{{\ld1\over m}\cos^2\th&\left[\cp+
(1-\GG)\sp\right]\cr&\qquad\qquad-\sin^2\th\left[\sp+\bar P\Z2\left({(m-1)
\over m}\cp+\sp\right)\right]\Biggr\},\cr}\eqno(2.35a)$$
$${\dd\ph\over\dd\ta}={1\over\st}\Biggl\{{\ld1\over m}\cos^2\th\left((1-\g1^{\
2})\cp-\sp\right)+\sin^2\th\;\bar P\Z2\left[{(m-1)\over
m}\sp-\cp\right]\Biggr\}
,\eqno(2.35b)$$
where
$$\bar P\Z2={m\over1+(m-1)\GG}\left\{\MgB1\cos^2\ph-2m\cp\sp+(m-1)\sin^2
\ph\right\}\,.\eqno(2.35c)$$
Four sets of critical points are found: \case (i):
Once again we obtain the endpoints of the curves of critical points with
$\lb=W=Z=0$. These points, which we will denote $L_{5-8}$ are located at
$$\th={\pi\over2},\qquad\ph=\arctan\left(m\pm\sqrt{1+\GG}\over m-1
\right)\,,\eqno(2.36a)$$
$$\hbox{or}\qquad\qquad X=\pm\infty,\qquad Y=X\left(m\pm\sqrt{1+\GG}\over
m-1\right),\qquad Z=0.\eqno(2.36b)$$
These points have the same properties as those outlined above for the
appropriate cases of the $\lb=W=Z=0$ points at finite distances from the
origin.
\case (ii): We once again obtain the critical points $M_1$ and $M_2$ given by
(2.29). As before, these points, corresponding to asymptotically flat
solutions, are found to act as saddles with respect to all trajectories other
than the 2-dimensional bunch of $W=Z=0$ Schwarzschild solutions.
\case (iii): If $\ld1>0$ and $\GG\le m+1$ or if $\ld1<0$ and $\GG\ge m+1$
then there are two critical points, which we will denote $N_1$ and $N_2$, which
are located at
$$\th=\arctan\left[\ld1\left({\Mg1\over m}+{m\over\Mg1}\right)\right]^{1/2},
\qquad\ph=\arctan\left(\Mg1\over m\right)\eqno(2.37a)$$
$$\hbox{or}\qquad\qquad X=\pm\infty,\qquad Y=\left(\Mg1\over m\right)X,\qquad
Z=\left(\Mg1\over m\ld1\right)^{1/2}X,\eqno(2.37b)$$
on the $\lb=0$ surface. In the full 5-dimensional phase space these points also
have
$$W=0,\qquad V=\left(\MggB12\over m\right)X.\eqno(2.37c)$$

If $\GG\le1$ then the point $N_1$ ($N_2$) in the first (third) quadrant
attracts (repels) a 3-dimensional set of trajectories in the 3-dimensional
subspace. ($N_1$ attracts all trajectories in the $\lb>0$ region, and some
$\lb<0$ trajectories if $\GG<1$.) If $1<\GG\le m+1$ then $N_1$ ($N_2$) only
attracts (repels) the 2-dimensional set of trajectories lying in the $\lb=0$
surface, and acts as a saddle with respect to other  directions. If $\GG=m+1$
then $N_1$ and $N_2$ are degenerate with the points $L_5$ and $L_7$. If $\GG>m+
1$ then the point $N_1$ ($N_2$) lying in the fourth (second) quadrant repels
(attracts) a 3-dimensional set of trajectories in the subspace.
\case (iv): If $\ld1>0$ then there are two critical points, which we will
denote $P_1$ and $P_2$, which are located at
$$\th=\arctan\left({2\ld1\over m}\mgB1\right)^{1/2},\qquad\ph={\pi\over4},\
{5\pi\over4}\eqno(2.38a)$$
$$\hbox{or}\qquad\qquad X=\pm\infty,\qquad Y=X,\qquad Z=\left(m\over\ld1\mgB1
\right)^{1/2}X.\eqno(2.38b)$$
In the full 5-dimensional phase space these points also have
$$W=0,\qquad V=\left(m\GG-\g1\g2+1\over\mg1\right)X.\eqno(2.38c)$$
If $\GG<1$ these points lie in the $\lb<0$ portion of the phase space, while if
$\GG>1$ they have $\lb>0$. If $\GG=1$ they are degenerate with points $N_1$ and
$N_2$.

If $\GG<1$ the point $P_1$ ($P_2$) in the first (third) quadrant attracts
(repels) a 2-dimensional set of $\lb<0$ trajectories, acting as a saddle with
respect to other directions. The 2-dimensional separatrix separates $\lb<0$
trajectories with an endpoint on $N_1$ ($N_2$) from trajectories with two
endpoints on the \ccurve. If $\GG\ge1$ then $P_1$ ($P_2$) attracts (repels) a
3-dimensional set of trajectories: all $\lb>0$ solutions apart from those lying
in the $Z=0$ plane.\smallskip

In Figs.~6 and 7 we sketch trajectories on the hemisphere at infinity of the
$W=0$ subspace for the various cases which give distinct behaviours.
\smallskip

The properties of the $Z=0$ subspace follow by symmetry upon making the
substitutions $Y\darr V$, $Z\darr W$, $\ld1\rarr\ld2$ and $\g1\darr\g2$ in the
above discussion. To set our notation, the critical points on the sphere at
infinity located at
$$\eqalign{X=\pm\infty,\qquad&V=\left(\Mg2\over m\right)X,\qquad W=\left(\Mg2
\over m\ld2\right)^{1/2}X,\cr&Z=0,\qquad Y=\left(\MggB12\over m\right)X,\cr}
\eqno(2.39)$$
will be denoted $R_1$ and $R_2$. The points located at
$$\eqalign{X=\pm\infty,\qquad&V=X,\qquad W=\left(m\over\ld2\mgB2\right)^{1/2}X,
\cr&Z=0,\qquad Y=\left(m\GGG-\g1\g2+1\over\mg2\right)X\cr}.\eqno(2.40)$$
will be denoted $Q_1$ and $Q_2$.
\subsection 2.4 Global properties of solutions

We turn now to the global properties of the solutions as deduced from the
nature of trajectories in the full 5-dimensional phase space with $\LA=0$. The
behaviour of the trajectories may be pieced together in a relatively
straightforward manner from the properties of trajectories in the subspaces
already discussed, since if $\LA=0$ the only critical points at infinity other
than those already found are the extension of points $L_{1-8}$ to the
one-parameter family of critical points which coincide with the intersection of
the $\lb=0$, $Z=0$, $W=0$ surface and the sphere at infinity. We shall denote
the whole set $\left\{L(y)\right\}$, where
$${m-\sqrt{\mg1}\over m-1}\le y\le{m+\sqrt{\mg1}\over m-1}\,.\eqno(2.41)$$
The points $L(y)$ are located at
$$\eqalign{&X=\pm\infty,\qquad Y=yX,\qquad Z=0,\qquad W=0,\cr&V=\left(m\g1(\gd)
+\mggB12y\pm\sqrt{2my-(m-1)y^2+\g1^{\ 2}-m-1}\over\mg1\right)X.\cr}\eqno(2.42)
$$

Generically, apart from a few exceptions\foot{$^{\hash7}$}{The $W=0$, $\lb=0$,
$\GG>1$ solutions given by equations (B.29)-(B.35) in Appendix B have endpoints
on $N_1$ and $N_2$ on the sphere at infinity, and have no asymptotic region. A
similar class of solutions exists for $Z=0$. Also, in the case of the
Robinson-Bertotti type solutions (2.20), some trajectories join the points $S_1
$ and $S_2$ (c.f.\ Fig.~2(b)).} trajectories which are not confined to the
sphere at infinity have at least one critical point on the \ccurve. Provided
that $\GG\le m+1$ and $\GGG\le m+1$, then all such critical points take
values $X_0$, $Y_0$ and $V_0$ which are either all positive, or all negative.
They respectively either repel or attract a 3-dimensional set of integral
curves in the 5-dimensional phase space, the remaining two directions being
degenerate. The points all correspond to $\rH=0$ singularities with the
exception of the points with $Y_0=V_0$ which correspond to horizons.

If $\GG>m+1$ or $\GGG>m+1$ then in addition to the critical points for
which $X_0$, $Y_0$ and $V_0$ are all of the same sign, critical points
of mixed signs also exist. These critical points are saddle points with respect
to most trajectories in the phase space. The $W=Z=0$ solutions form one
separatrix of trajectories with endpoints at the saddle points; for these
solutions we still have $\rH\rarr0$ as the critical points are approached.
Other separatrices are formed by the $\lb=W=0$ solutions, or the $\lb=Z=0$
solutions, as appropriate. For these solutions $\rH\rarr\infty$ as the saddle
points are approached, as is discussed in Appendix B. For most trajectories,
however, including the $\lb>0$ ones which are of prime interest to us, it is
necessary to examine the behaviour of the solutions at the phase space infinity
in order to determine the asymptotic ($\rH\rarr\infty$) behaviour of solutions
for which an asymptotic region exists. Apart from such solutions there are also
many trajectories with two endpoints on the \ccurve\ which have no asymptotic
region. Generally, they connect two points at which $\rH\rarr0$. However, a
subset with two endpoints in the $V=Y$, $W=\ga Z$ subspace describe
Schwarzschild-de Sitter like solutions, as was discussed in \S2.2.

All critical points at infinity other than the points $L(y)$ and $S_{1,2}$ are
found to correspond to $\rH\rarr\infty$ provided that $0<\GG<m+1$ and $0<\GGG<
m+1$. If $\g1=0$ then the points $P_{1,2}$ correspond to $\rH\rarr\const$
(indicating the presence of Robinson-Bertotti-like solutions), while the points
$N_{1,2}$ still correspond to $\rH\rarr\infty$. The same is true for the points
$Q_{1,2}$ and $R_{1,2}$ respectively if $\g2=0$. If $\GG>m+1$ then points $N_{1
,2}$ correspond to $\rH\rarr0$, while if $\GGG>m+1$ then the points $R_{1,2}$
similarly correspond to $\rH\rarr0$. To discuss the asymptotic form of the
solutions it is perhaps more convenient to examine the behaviour of the metric
functions of the more usual Schwarzschild-type coordinates
$$\gxxh=-\euu\dd t^2+\evv\dd\rH^2+\rH^2\gxxb,\eqno(2.43)$$
rather than (2.1a). In Table 1 we display the asymptotic form of the metric
functions (2.43), and of the scalar field, for integral curves from regions
of the phase space at a finite distance from the origin which approach each of
the critical points on the sphere at infinity for which $\rH\rarr\infty\;$\foot
{$^{\hash8}$}{If $\GG>m+1$ or $\GGG>m+1$ then we will also have $\rH\rarr
\infty$ for $\lb=0$ trajectories which approach points $L(y)$ with $y<0$.
However, such trajectories are confined to the sphere at infinity and thus do
not represent physical integral curves, and so we omit them.}.
\midinsert \def\spil{height2pt&\omit&&\omit&&\omit&&\omit&&\omit&\cr}
\def\ar{\cr\spil\noalign{\hrule}\spil}
$$\vbox{\offinterlineskip\hrule\halign{&\vrule#&\strut\quad$\dsp#$\quad\hfil\cr
\spil&\omit&&\hbox{Values of constants}&&\euu&&\evv&&e^{2\ka\si}&\ar
&M_{1,2}&&\lb>0&&\const&&\const&&\const&\ar&N_{1,2}&&\GG<m+1,\ \ld1>0,\ \lb=0&&
\rH^2&&\rH^{2(\GG-1)}&&\rH^{m\g1}&\ar&P_{1,2}&&\g1\ne0,\ \ld1>0,\ \hbox{sign}\;
\lb=\hbox{sign}\;(\GG-1)&&\rH^{2/\GG}&&\const&&\rH^{m/\g1}&\ar&Q_{1,2}&&\g2\ne0
,\ \ld2>0,\ \hbox{sign}\;\lb=\hbox{sign}\;(\GGG-1)&&\rH^{2/\GGG}&&\const&&\rH^{
m/\g2}&\ar &R_{1,2}&&\GGG<m+1,\ \ld2>0,\ \lb=0&&\rH^2&&\rH^{2(\GGG-1)}
&&\rH^{m\g2}&\ar &T_{1,2}&&\LA<0,\ \lb=0&&\rH^2&&\rH^{-2}&&\const&\cr \spil}
\hrule}$$
\caption{Table 1}{Asymptotic form of solutions for trajectories approaching
critical points at infinity from within the sphere at infinity.}
\endinsert

In order to classify the various solutions we must first of all determine the
nature of the various critical points at infinity. It is straightforward but
laborious to evaluate the eigenvalue spectrum for small perturbations about the
points. In Table 2 we summarise the results for such an analysis. We display
the eigenvalues for the points with $X>0$. For the corresponding points with
$X<0$ the sign of the eigenvalues is simply reversed.

The qualitative behaviour of the trajectories is largely dependent on the
values of $\g1$ and $\g2$, apart from the case of points $M_{1,2}$. These
points -- the only ones which correspond to solutions asymptotically flat in
terms of $\rH$ -- are endpoints for a 3-dimensional set of solutions for all
values of $\g1$ and $\g2$. These solutions are just those lying in the $W=Z=0$
subspace, which is physically equivalent to $\ld1=\ld2=\dots=\ld s=0$. Thus
models with non-zero $\ld i$ possess no solutions which are asymptotically
flat in terms of $\rH$ if $\LA\ne0$.

The eigenvalues for small perturbations near the points $N_{1,2}$, $P_{1,2}$,
$Q_{1,2}$ and $R_{1,2}$ are essentially independent of the constants $\g i$,
$i\ge3$. The only exceptions is one eigenvalue at each point in the case that
at least one of the $\al_i$ or $\be_i$ is less than one half. On account of the
ordering (2.2) we also have
$$0<\al\Z3<\al_4<\dots<\al_{s-1}<\al_s<1,\eqno(2.44)$$
and
$$1>\be_3>\be_4>\dots>\be_{s-1}>\be_s>0.\eqno(2.45)$$
\topinsert \def\spil{height2pt&\omit&&\omit&\cr}
\def\ar{\cr\spil\noalign{\hrule}\spil}
$$\vbox{\offinterlineskip\hrule\halign{&\vrule#&\strut\ $\dsp#$\ \hfil\cr
\spil&\omit&&\hbox to16true mm{\hfil}\hbox{Eigenvalues (with degeneracies)}&\ar
&L(y)&& 0,\ ({\it2});\ 2;\ y;\ v.^*&\ar
&M_1&&-1,\ ({\it3});\ {1\over m-1},\  ({\it2}).&\ar
&N_1&&{-1\over m}\MgB1,\ ({\it3});\ {2\over m}\left(\GG-1\right);\ {\g1
\over m}\left(\g1-\g\al\right).&\ar
&P_1&&-1,\ ({\it2});\ {-1\over2}\left[1\pm\sqrt{9+(m-9)\GG\over\mg1}\,
\right];\ {\g1\left(\g1-\g\al\right)\over\mg1}\,.&\ar
&Q_1&&-1,\ ({\it2});\ {-1\over2}\left[1\pm\sqrt{9+(m-9)\GGG\over\mg2}
\,\right];\ {-\g2\left(\g\be-\g2\right)\over\mg2}\,.&\ar
&R_1&&{-1\over m}\MgB2,\ ({\it3});\ {2\over m}\left(\GGG-1\right);\
{-\g2\over m}\left(\g\be-\g2\right).&\ar
&S_1&&-2;\ -1;\ 1; {1\over2m}\left\{-m\pm\sqrt{m^2-{8m(\gd)\lag\over\LA}}\,
\right\}.&\ar
&T_1&&{-(m+1)\over m},\ ({\it2});\ {-2\over m};&\cr\spil\spil &\omit&&\
{1\over2 m}\left\{-m-1\pm\sqrt{(m+1)^2-{8(m+1)(\gd)\lag\over\LA}}\,\right\}.&
\cr\spil}\hrule}$$
\caption{Table 2}{Eigenvalues of critical points at infinity. The eigenvalues
for small perturbations which are degenerate have the degeneracy listed in
brackets.\br $^*$ The values of $y$ and $v$ listed are defined by (2.41) and
$V=vX$ in (2.42).}
\endinsert
\noindent Thus either $\al\Z3$ or $\be_s$ has the smallest value of the $\al_i$
and $\be_i$ -- this value being important in defining coordinates at the phase
space infinity which lead to a well-defined spectrum of linearised
perturbations. If we define
$$\g\al=\cases{\g3,&if $\al\Z3<\be_s$,\cr \g2,&otherwise,\cr}\eqno(2.46)$$
and
$$\g\be=\cases{\g s,&if $\be_s<\al\Z3$,\cr \g2,&otherwise,\cr}\eqno(2.47)$$
then we find that one of the eigenvalues at the points $N_{1,2}$, $P_{1,2}$
depends on the factor $(\g1-\g\al)$, while one of the eigenvalues at the
points $Q_{1,2}$, $R_{1,2}$ depends on the factor $(\g\be-\g2)$. However, since
both these factors are positive for each choice of $\al$ and $\be$ there is no
qualitative difference between the alternatives.

In Table 3 we summarise the nature of the set of solutions with endpoints at
$N_{1,2}$ and $P_{1,2}$. The corresponding results for $R_{1,2}$ and $Q_{1,2}$
respectively may be obtained by substituting $\g1\rarr\g2$, $W\rarr Z$.
We display the dimension, $d_\A$, of the maximal set, $\A$, of
trajectories with endpoints at each point -- for $\GG\le m+1$ this means the
dimension of the set of trajectories attracted to (repelled from) the point
$N_1$ ($N_2$), and vice-versa if $\GG>m+1$. In the case of the point $P_1$
($P_2$) it means the dimension of the set of trajectories attracted (repelled).
Each point with $d_\A<5$ will also be the endpoint for a $(5-d_\A)$-dimensional
separatrix of saddle point trajectories.
\midinsert
\def\spil{height2pt&\omit&\omit&\omit&&\omit&&\omit&&\omit&&\omit&\cr}
\def\ar{\cr\spil\noalign{\hrule}\spil}
\def\aar{\cr\spil&\omit&\multispan{11}\hrulefill\cr\spil}
$$\vbox{\offinterlineskip\hrule\halign{&\vrule#&\strut\ $\dsp#$\ \hfil\cr
\spil&\multispan3\hfil$\dsp N_{1,2}$\hfil&&\GG<1&&\GG=1&&1<\GG\le m+1&&\GG>m+1
&\cr\spil\noalign{\hrule}\spil &\omit&\omit&d_\A&&4&&4&&3&&5&\aar &\g1\ge0&
\omit&\hbox{Nature of}&&W=0,\ \lb\ge0\,;&&W=0,\ \lb\ge 0\,;&&W=0,\ \lb=0&&\lb
\ge0\,;&\cr \spil &\omit&\omit&\hbox{solutions}&&\omit&&\hbox{$3$-dim}\ $W=0$,&
&\omit&&\omit&\cr \spil&\omit&\omit&\{\A\}&&W=0,\ \lb<0&&\ \lb<0\ \hbox
{separatrix}&&\omit&&\lb<0&\ar &\omit&\omit&d_\A&&5&&5&&4&&4&\aar &\g1<0&\omit&
\hbox{Nature of}&&\lb\ge0\,;&&\lb\ge0\,;&&\lb=0&&W=0,\ \lb\ge0\,;&\cr \spil
&\omit&\omit&\hbox{solutions}&&\omit&&\hbox{$4$-dim}\ \lb<0&&\omit&&\omit&\cr
\spil&\omit&\omit&\{\A\}&&\lb<0&&\ \hbox{separatrix}&&\omit&&W=0,\ \lb<0&\cr
\spil}\hrule}$$
\def\spil{height2pt&\omit&\omit&\omit&&\omit&&\omit&\cr}
\def\ar{\cr\spil\noalign{\hrule}\spil}
\def\aar{\cr\spil&\omit&\multispan7\hrulefill\cr\spil}
$$\vbox{\offinterlineskip\hrule\halign{&\vrule#&\strut\ $\dsp#$\ \hfil\cr
\spil&\multispan3\hfil$\dsp P_{1,2}$\hfil&&\GG<1&&\GG>1&\cr
\spil\noalign{\hrule}\spil &\omit&\omit&d_\A&&3&&4&\aar &\g1\ge0&\omit&\hbox
{Nature of}&&W=0,\ \lb<0&&W=0,\ \lb>0&\cr \spil &\omit&\omit&\hbox{solutions}\
\{\A\}&&\hbox{separatrix}&&\omit&\ar &\omit&\omit&d_\A&&4&&5&\aar &\g1<0&\omit&
\hbox{Nature of}&&\lb<0&&\lb>0&\cr \spil&\omit&\omit&\hbox{solutions}\ \{\A\}&&
\hbox{separatrix}&&\omit&\cr\spil}\hrule}$$
\caption{Table 3}{Nature of trajectories that approach points $N_{1,2}$ and
$P_{1,2}$.}
\endinsert
\noindent Of most interest are the points which are endpoints for a
5-dimensional set of trajectories, as they represent solutions with the most
typical behaviour. If $\g1>0$ then such points ($N_{1,2}$) exist only when
$\GG>m+1$ and $\ld1<0$. Such points exist for all $\g1<0$ if $\ld1>0$: for
$\GG\le1$ they are the points $N_{1,2}$, and for $\GG>1$ the points $P_{1,2}$.
For the higher derivative theories, however, $\g1>0$ and $\GG<m+1$ and thus
the points $N_{1,2}$ or $P_{1,2}$ are endpoints for at most a 4-dimensional
set of solutions.

The nature of the set of solutions with endpoints at $S_{1,2}$ and $T_{1,2}$ is
dependent on both the constants $\g i$ and the constants $\ld i$. However, the
three eigenvalues corresponding to directions which lie within the (anti)-de
Sitter subspace $V=Y$, $W=\ga Z$ are independent of the $\g i$ and $\ld i$, and
so any differences are determined by the remaining two eigenvalues, which are
given by the solutions of the equations
$$\la\Z S^{\ 2}+\la\Z S+{2(\gd)\lag\over m\LA}=0\eqno(2.48a)$$
for the points $S_{1,2}$, and
$$\la\Z T^{\ 2}+\left(m+1\over m\right)\la\Z T+{2(m+1)(\gd)\lag\over m^2\LA}=0
\eqno(2.48b)$$
for the points $T_{1,2}$, where
$$\lag=\ld2\g 2\ga^2+\ssum\ld i\g i\al\Z i\ga^{2\al_i}=-\ld1\g1-\ssum\ld i\g i
\be\Z i\ga^{2\al_i}.\eqno(2.48c)$$
For all choices we find only two possibilities: the dimension of the maximal
set of trajectories with endpoints at $S_{1,2}$ is either three or four; while
the dimension of the maximal set of trajectories with endpoints at $T_{1,2}$ is
either four or five. The later was true in the case of the \KK\ models studied
in II.

This completes our classification of the solutions for Einstein gravity coupled
to a scalar field with potential (1.10) for which the constant $\LA$, defined
by (2.14), (2.15d) is non-zero. For non-zero $\ld i$ solutions which are
asymptotically flat in terms of $\rH$ do not exist. This is not the case if
$\LA=0$, however, as we shall see in the next section.
\section 3. Solutions with a Schwarzschild subspace (Special polynomial $R$
theories)

If there exist solutions $\ga$ of (2.14) such that the constant $\LA$ defined
by (2.15d) vanishes, then the structure of the phase space is significantly
altered. No such solutions exist if $s=2$. However, if $s\ge3$, which applies
in the case of the \rsq\ and other higher derivative models with potential
(1.13), then such solutions are possible. We see from (2.17) that if $\LA=0$
we immediately obtain the Schwarzschild solution -- thus all solutions lying in
the 3-dimensional subspace $V=Y$, $W=\ga Z$ (for appropriate $\ga$), and not
just those in the $Z=0$ plane, are Schwarzschild solutions. This may be
verified by direct integration since equations (2.15) are now equivalent to the
equations which lead to the solutions of Appendix B.1, with the added
restriction that $\cd1=mk$.

If we compare the phase space structure with that of the subspace of \S2.2, we
observe that the points $S_{1,2}$ and $T_{1,2}$ do not exist in the
Schwarzschild subspace. Instead, we have an additional one parameter family of
critical points at a finite distance from the origin. These may be
parameterised in terms of their arbitrary value of $Z=Z_0>0$, and are located
at
$$X=Y=V=0,\qquad Z=Z_0,\qquad W=\ga Z_0.\eqno(3.1)$$
We shall denote these points $\OZ$, and the point at infinity ($Z_0\rarr
\infty$) will be denoted $O_1$. These are the only additional critical points.

Small perturbations about the points $\OZ$ yield three zero eigenvalues,
indicating a high degree of degeneracy. However, their properties with regard
to the structure of the phase space may be ascertained from the following
observations. Firstly, if $Y=X$ or $Y=(m+1)X/(m-1)$, then $P=0$, $X'=0$ and
$Y'=0$, so that the motion of the trajectories is entirely in the $Z$
direction, as is depicted in Fig.~8. Furthermore, since (2.23) is now true
for all solutions we may foliate the 3-dimensional subspace by a stack of
planes $Y=m(X+k)/(m-1)$ to which trajectories are confined. Each plane is
described by a 2-dimensional autonomous system
$$X'=X^2-m^2k^2,\eqno(3.2a)$$
$$Z'={m\over m-1}\left(X+k\right)Z,\eqno(3.2b)$$
with solution
$$Z=C_0\left|X+mk\right|^{1/2}\left|X-mk\right|^{(m+1)/[2(m-1)]}\eqno(3.3a)$$
if $k\ne0$, or
$$Z=C_0X^{m/(m-1)},\eqno(3.3b)$$
if $k=0$, where $C_0$ is an arbitrary constant in both cases. The pattern of
these trajectories is depicted in Fig.~9. The critical points at $\pm mk$ are
of course those that lie on the \ccurve, and as expected they either attract or
repel a 2-dimensional bunch of solutions in the subspace.  From Figs.~8 and 9
we can see that for finite $Z_0$ the points $\OZ$ neither attract nor repel
any trajectories, and so there are no solutions for which they are endpoints.
Clearly, however, all trajectories which reach infinity other than those in the
planes depicted in Fig.~8 approach the point $O_1$. This is borne out by the
plot of the hemisphere at infinity with coordinates $\th$ and $\ph$ defined by
(2.24), which is shown in Fig.~10(a).

The fact that $O_1$ appears to be an endpoint for many solutions is in fact
something of a misnomer, which arises from the fact that $Z\goesas X^{m/(m-1)}$
as $X\rarr\infty$ for all solutions (3.4) with an asymptotic region, so that
$Z$ grows more rapidly than either $X$ or $Y$. Thus although the solutions
(3.3) do not have $X=0$ or $Y=0$ as $Z\rarr\infty$, they are nonetheless
projected onto the north pole, $O_1$, if coordinates (2.24) are used. This
degeneracy can be lifted if instead of (2.24) we use coordinates $(\rh,\th,\ph)
$ defined by
$$X=\rh\st\cp\,\qquad Y=\rh\st\sp\,\qquad Z^{(m-1)/m}=\rh\ct\,,\eqno(3.4)$$
and perform the same analysis as before. This yields the plot Fig.~10(b) for
the sphere at infinity. In these coordinates we obtain a line of critical
points on the sphere at infinity with arbitrary angle $\th$,
$$0<\th<{\pi\over2},\qquad\ph=\arctan\left(m\over m-1\right)\,.\eqno(3.5)$$
Perturbations about these critical points still yield one zero eigenvalue,
corresponding to the degenerate direction $\th$, but each point in the
first (third) quadrant is found to attract (repel) a 2-dimensional set of
solutions from regions of the phase space at a finite distance from the origin.

The crucial question now is: do any trajectories which lie outside the
Schwarzschild subspace have an endpoint at $O_1$? If such trajectories do
exist, and if any of them curve back to another endpoint in the Schwarzschild
subspace corresponding to an horizon, then we would have solutions with
non-trivial scalar fields which violate the no-hair theorems. The corresponding
Schwarzschild solution in any equivalent higher derivative theory would be
non-unique. The answer is that the eigenvalues for perturbations in the two
extra directions are given by
$$\la_O^{\ 2}={-2\over m}(\gd)\lag,\eqno(3.6)$$
where $\lag$ is given by (2.48c). Provided that $\lag>0$ then $O_1$ will be a
``centre'' with respect to the extra directions, and the only solutions with
an endpoint there will indeed be just the Schwarzschild solutions. If $\lag<0$,
however, then $O_1$ will be a saddle point with respect to the extra two
directions, and a 4-dimensional separatrix of solutions will have an endpoint
at $O_1$ within the full 5-dimensional phase space. Unfortunately, it is not
immediately obvious whether the solutions with an endpoint at $O_1$ which lie
outside the Schwarzschild subspace have a second endpoint on the \ccurve\ at a
point corresponding to an horizon or to a singularity. Thus we cannot extend
the no-hair theorem, or the corresponding uniqueness theorem for higher
derivative theories, to the case in which $\lag<0$.

For the $R+a R^2$ theory, or indeed for any action (1.7) with coefficients
(1.11), which gives rise to an equivalent potential (1.13), we find that
a Schwarzschild subspace is obtained only for $\tee=1$ and $\ga=1$.
Consequently, $\lag=m/(8a\sqrt{m+1})$ and
$$\la\Z O^{\ 2}={-m\over4a(m+1)}\,,\eqno(3.7)$$
so that $O_1$ is an endpoint for Schwarzschild solutions only provided $a>0$.
This is of course the same condition required for Whitt's proof in the case of
the \rsq\ (c.f.\  Appendix A).

To complete our proof of the uniqueness theorem for higher derivative black
holes it is still necessary to check the asymptotic behaviour of the metric
functions near the critical points at infinity for which $e^{2\ka\si}$ is not
asymptotically constant in terms of the physical radial coordinate $r$ instead
of the coordinate $\rH$ used in Table 1. The results for higher derivative
actions with equivalent potential (1.13) are listed in Table 4, in terms of
metric functions for the physical metric in Schwarzschild-type coordinates:
$$\exp\left(-4\ka\si\over m\sqrt{m+1}\right)\gxxh=-e^{2{\rm u}}\dd t^2+e^{2{\rm
v}}\dd r^2+r^2\gxxb.\eqno(3.8)$$
\midinsert \def\spil{height2pt&\omit&&\omit&&\omit&&\omit&\cr}
\def\ar{\cr\spil\noalign{\hrule}\spil}
$$\vbox{\offinterlineskip\hrule\halign{&\vrule#&\strut\ $\dsp#$\ \hfil\cr
\spil&\omit&&\hbox{Values of constants}&&e^{2{\rm u}}&&e^{2{\rm v}}&\ar
&N_{1,2}&&{\tee^{n-1}\over a}<0,\ \lb=0&&r^2&&r^m&\ar &P_{1,2}&&{\tee^{n-1}
\over a}<0,\ \lb>0&&r^{4/(m+2)}&&\const&\ar &Q_{1,2}&&m\ne2(n-1),\ {\tee\over a
}<0,&&r^{4(n-1)(2n-1)/[m-2(n-1)]}&&\const&\cr\spil &\omit&&\hbox{sign}\;\lb=
\hbox{sign}\;[m-4n(n-1)]&&\omit&&\omit&\ar &R_{1,2}&&{\tee\over a}<0,\ \lb=0
&&r^2&&r^{[m-4n(n-1)]/[(n-1)(2n-1)]}&\cr \spil}\hrule}$$
\caption{Table 4}{Asymptotic form of the higher derivative theory metric
for polynomial $R$ actions (1.7)  with coefficients (1.11) in terms of the
physical radial coordinate $r$, for the critical points at infinity at which
$r$ is not proportional to $\rH$.}
\endinsert
\noindent (Near the points $T_{1,2}$ the solutions are of course still
asymptotically anti-de Sitter in terms of the coordinate $r$ as well as $\rH$.)
It is evident that, as required, none of these solutions are asymptotically
flat.

{}From Table 3, we see that for the higher derivative theories points $N_{1,2}$
are endpoints for the 3-dimensional set of $W=0$, $\lb=0$ solutions, while
points $P_{1,2}$ are endpoints for the 4-dimensional set of $W=0$, $\lb>0$
solutions. The constant $\g2$ can take either sign depending on the relative
values of $m$ and $n$, and its absolute value is only restricted by $\GGG<m+1$.
Thus the dimension of the maximal set of solutions with endpoints at $R_{1,2}$
and $Q_{1,2}$ is three, four or five depending on the particular values of $m$
and $n$. For $m=2$ (i.e.\ the 4-dimensional theory) and $n\ge3$, points $R_{1,2
}$ are endpoints for a 5-dimensional set of solutions of either sign of $\lb$,
while points $Q_{1,2}$ are similarly endpoints for a 4-dimensional separatrix.
If $m=2$ and $n=2$, the dimension of these sets is reduced by one with the
additional restriction that $Z=0$. In this case -- or for any $m$ and $n$ such
that $m=2(n-1)$ -- the points $R_{1,2}$ correspond to asymptotically anti-de
Sitter solutions. These solutions appear to form a special class unrelated to
those of \S2.2.

The points $S_{1,2}$ and $T_{1,2}$ exist provided that $m\ne2(n-1)$, $\ga$
being given by roots of the polynomial (2.14):
$$[2(n-1)-m]\teega^n-2n\teega^{n-1}+m+2=0,\eqno(3.9)$$
for which $\tee\ga^{2/n}\ne1$, (and $\ga>0$). If we factor out the
Schwarzschild root (3.9) reduces to the polynomial
$$[2(n-1)-m]\teega^{n-1}-(m+2)\sum_{j=0}^{n-2}\teega^j=0.\eqno(3.10)$$
For the \rsq\ ($n=2$), for example, the only solution (with $\ga>0$) is
$$\tee=-1,\qquad\ga=\left(m+2\over m-2\right),\qquad m\ge2,\eqno(3.11)$$
which gives
$$\ga^{-1}\LA={-m\over(m+2)(m-2)a}\,,\eqno(3.12)$$
as is appropriate for the exact solution (2.19). Thus the points $S_{1,2}$
and $T_{1,2}$ exist only for $a<0$, $m>2$. We find that
$$\lag={-m\over8a\sqm}\left(m+2\over m-2\right)\,\eqno(3.13)$$
and consequently a 4-dimensional set of trajectories have endpoints at $S_{1,2
}$, and a 5-dimensional set of trajectories have endpoints at the points $T_{1,
2}$, with anti-de Sitter asymptotics. If $a>0$, the structure of the phase
space is completely different: the points $S_{1,2}$ and $T_{1,2}$ do not exist.
Instead we find a class of solutions with two endpoints lying in the $V=Y$,
$W=(m+2)Z/(m-2)$ subspace -- these are asymptotically de-Sitter. In the case of
four dimensions $(m=2)$ no subspaces of purely Schwarzschild-(anti) de Sitter
solutions exist in the \rsq. However, the points $R_{1,2}$ (which are endpoints
for a 4-dimensional set of solutions) nonetheless have anti-de Sitter
asymptotics.
\section 4. General polynomial $R$ theories

The analysis of the previous two sections applies exactly only to those
polynomial $R$ theories with coefficients (1.11). For lagrangians of degree
$n\ge3$ in $R$ we shall see, however, that in the general case the field
equations become equivalent to those given by one exponential sum potential or
another near all the critical points. Consequently, the global properties of
all theories polynomial in $R$ can be obtained from the results of \S2 and \S3.
As an independent check on this argument in one case, however, we shall first
treat the general cubic theory in detail.

\subsection 4.1 The $R+aR^2+bR^3$ theory

The general cubic theory with $f=R+aR^2+bR^3$ is described by a potential (1.6)
which contains terms non-polynomial in $\exp\left[\ka\si/\sqrt{D-1}\right]$.
The potential (1.6) may be conveniently written as
$$\eqalign{\V={-1\over4\ka^2}\Biggl\{\ld1\exp\left(-4\g1\ka\si\over m\right)&+
\ld2\exp\left(-4\g2\ka\si\over m\right)\left[{b\ee\over|b\ee|}+{\ld4\over\ld2}
\exp\left(-8(\gd)\ka\si\over3m\right)\right]\cr&\qquad\qquad+\ld3\exp\left(-4
\g3\ka\si\over m\right)\Biggr\}\cr},\eqno(4.1a)$$
where
$$\g 1={m+2\over2\sqm},\qquad\g 2={-(m-4)\over4\sqm},\qquad\g 3={1\over\sqm},
\eqno(4.1b)$$
$$\ld1={2a\ee\over27b^2}\left(a^2-{9\over2}b\right),\qquad\ld2={-2\ee\tee a
\over3\sqrt{3}\;|a||b|^{1/2}},\qquad\ld3={a\over3b},\qquad\ld4={\ld2\over3b}
\left(a^2-3b\right),\eqno(4.1c)$$
and $\tee$ is now defined by
$$\tee=\cases{1,&if $1+3bR/a>0$,\cr -1,&if $1+3bR/a<0$.\cr}\eqno(4.1d)$$
As before, $\ee$ is defined by (1.2b). The quantities $\g1$, $\g2$ and $\g3$
are still identical to those given by (2.10a) for $n=3$. However, the
coefficients $\ld i$ now differ in general from those given by (2.10c) -- the
new coefficients $\ld i$ reduce to the values (2.10c), with $\ld4=0$, in the
special case $a^2=3b$, when the potential is an exponential sum. Our new
definition for $\tee$ also reduces to the former definition (1.13b) if $a^2=3b
$. The Einstein-scalar field equations with the potential (4.1) are given by
$$\ze''=(m-1)^2\lbz+\laet+\lacch+\lcross,\eqno(4.2a)$$
$$\eqalign{\et''=\ &m(m-1)\lbz+{1\over m}\MgB1\laet\cr&+{1\over m}\MggB12\lacch
\!+{1\over m}\MggB13\lcross\cr &\qquad\qquad-{\g1\over m}\lccross,\cr}\eqno
(4.2b)$$
$$\eqalign{\ch''=\ &m(m-1)\lbz+{1\over m}\MggB12\laet\cr&+{1\over m}\MgB2\lacch
\!+{1\over m}\MggB23\lcross\cr&\qquad\qquad-{\g2\over m}\lccross,\cr}\eqno
(4.2c)$$
with the constraint
$$\eqalign{(m+1)\ze'^2&+{2m\ze'(\g 2\et'-\g 1\ch')\over\gd}+{\mgB2\over(\gd)^2}
{\et'}^2\cr&\ -2{\mggB12\over(\gd)^2}\et'\ch'+{\mgB1\over(\gd)^2}{\ch'}^2+(m-1)
\lbz\cr&\ \qquad+{\laet\over m}+{\lach\over m}e^{2\varch}\bech^{3/2}+{\ld3\over
m}e^{2(\et+2\varch)/3}=0,\cr}\eqno(4.2d)$$
where the variables $\ze$, $\et$ and $\ch$ are defined as before by
(2.3)-(2.5). If we now define the variables $X$, $Y$, $Z$, $V$ and $W$ by (2.7)
once again, and use the constraint (4.2d) to eliminate the $e^{2\ze}$
terms from (4.2a-c) we obtain the first-order system
$$X'={1\over m}\left\{\lz1+\ld2\GA^2+\lwzt\right\}-{(m-1)P\over m},\eqno(4.3a)
$$
$$\eqalign{Y'={-1\over m}\Biggl\{\gs1\lz1+\gg12\ld2\GA^2&+\gg13\lwzt\cr&+\g1
\lwzr\Biggr\}-P,\cr}\eqno(4.3b)$$
$$\eqalign{V'={-1\over m}\Biggl\{\gg12\lz1+\gs2\ld2\GA^2&+\gg23\lwzt\cr&+\g2
\lwzr\Biggr\}-P,\cr}\eqno(4.3c)$$
$$Z'=YZ,\eqno(4.3d)$$
$$W'=VW,\eqno(4.3e)$$
where
$$\GA\equiv\left[{b\ee\over|b\ee|}\,W^{4/3}+{\ld4\over\ld2}Z^{4/3}\right]^{3/4}
,\eqno(4.3f)$$
and $P$ is still given by (2.8f).

The differences between equations (4.3) and (2.8) do not give rise to any
significant changes to the analysis of \S2 and \S3. The position of all
critical points $W=0$ and $Z=0$, and in particular the $W=0$, $Z=0$ $\lb=0$
surface, is unchanged, as are the eigenvalues for small perturbations about
them. The number of critical points at infinity with $Z\ne0$ or $W\ne0$ is also
the same, and their location is more or less the same. For points $R_{1,2}$ and
$Q_{1,2}$ the definitions (2.39) and (2.40) are the same, if we now use $\ld2$
as defined in (4.1c). These points now exist if $\ee\tee/a<0$. In the case of
points $N_{1,2}$ and $P_{1,2}$, we must make the replacement $\ld1\rarr\ld1+\ld
4\left[\ld4/\ld2\right]^{1/2}$ in the definitions (2.37) and (2.38), with $\ld1
$, $\ld2$ and $\ld4$ given by (4.2c). These points now exist if
$$b>0,\qquad a^2\ge3b,\qquad\tee=+1,\qquad\ee a<0,\eqno(4.4)$$
or if
$$b>0,\qquad a^2\ge3b,\qquad\tee=-1,\qquad\left(a^2-4b\right)\ee a>0.\eqno(4.5)
$$
In the case of points $S_{1,2}$, $T_{1,2}$ and $O_1$, our previous
definitions are once again valid if we replace equations (2.14) and (2.15d),
which respectively define $\ga$ and $\LA$, by
$$\ld1\g1+\ld2\g2\Gag^{3/2}+\ld3\g3\ga^{4/3}+\ld4(\gd)\Gag^{1/2}=0,\eqno(4.6)$$
and
$$\LA={-1\over m}\left\{\ld1+\ld2\Gag^{3/2}+\ld3\ga^{4/3}\right\}\eqno(4.7)$$
Here the plus (minus) sign corresponds to $b\ee>0$ ($b\ee<0$).
The eigenvalues about the various critical points are unchanged from those
given in Table 2, if $\lag$ is defined by
$$\eqalign{\lag&=\ga^{4/3}\left\{\ld2\g2\Gag^{1/2}+{2\over3}\ld3\g3+{1\over3}
\ld4(\gd)\Gag^{-1/2}\right\}\cr&=-\ld1\g1-{1\over3}\ld3\g3\ga^{4/3}-\Gag^{-1/2}
\left[{1\over3}(2\g1+\g2)\ga^{4/3}+{\ld4\over\ld2}\g1\right]\,,\cr}\eqno(4.8)$$
instead of (2.48c).

A Schwarzschild subspace ($\LA=0$) is obtained only for $\ee=+1$ and $\tee=+1$,
the appropriate solution then being given by $\ga=1$. Eigenvalues for small
perturbations within the subspace give the same results as previously, and
furthermore we find that for perturbations in the additional two directions
the eigenvalues are once again
$$\la\Z O^{\ 2}={-m\over4a(m+1)}\,.\eqno(4.9)$$
Thus just as in the special case of polynomial actions with coefficients
(1.11), the black hole uniqueness theorem applies to solutions with $a>0$.
Most significantly, this result applies independently of the coefficient $b$ of
the $R^3$ term.

With regard to the (anti)-de Sitter subspaces, we find that (4.6) has the
solution
$$\ga^{4/3}=\cases{\dsp{(m+2)\left[(m-2)a^2-4(m-4)b\pm a\sqrt{(m-2)^2a^2-4m(m-4
)b}\right]\over2(m-4)^2b\ee}\,,& if $m\ne4$,\cr \dsp{3\left(4b-a^2\right)\over
a^2\ee}\,,& if $m=4$,\cr}\eqno(4.10)$$
where the sign of $\ee$ is taken so as to make the r.h.s.\ positive.
Consequently
$$\ga^{-4/3}\LA=\cases{\dsp{-(m-2)a\mp\sqrt{(m-2)^2a^2-4m(m-4)b}\over2(m+2)(m-4
)b}\,,& if $m\ne4$,\cr \dsp{-1\over3a}\,,& if $m=4$.\cr}\eqno(4.11)$$
Thus exact Schwarzschild-de Sitter-type solutions of the form (2.18) are also
obtained for $m=2$, in
contrast to the \rsq. Furthermore, although in the case of the \rsq\ only
anti-de Sitter solutions are admissable if $a>0$, for the present theory
both de Sitter and anti-de Sitter solutions are obtained for $a>0$ if $m\ne4$,
(for either sign of $b$).
\subsection 4.2 Theories of arbitrary degree

At the next order with $f=R+aR^2+bR^3+cR^4$ the polynomial giving $R$ in terms
of $f'$ has three branches, giving rise to potentials such as
$$\eqalign{\V={\ee\over4\ka^2c^3}\sigo\Biggl\{&{b^4\over256}+c^2q-2bc^3\SI\Z0
\cr&-6qc^2\left[\SI\Z++\SI\Z-\right]+3c^4\left[\SI\Z+^{\ 2}-\SI\Z-^{\ 2}\right]
\Biggr\},\cr}\eqno(4.12a)$$
where
$$q={1\over16c^2}\left({8\over3}ac-b^2\right),\eqno(4.12b)$$
$$\SI\Z0\equiv{1\over64c^3}\left[4abc-b^3-8c^2\ee\sig\right],\eqno(4.12c)$$
and
$$\SI\Z{\pm}\equiv\left(\SI\Z0\pm\sqrt{q^3+\SI\Z0^{\ 2}}\,\right)^{2/3}.\eqno
(4.12d)$$
Such a potential would lead once again to equations similar to (4.3), if the
$\g i$ of (2.10a) with $n=4$ are used, but now with a further level of nesting
of terms involving irrational roots. In the special case that $3b^2=8ac$,
(i.e., $q=0$), $\SI\Z+=2\SI\Z0$ and $\SI\Z-=0$, so that the potential (4.12)
and the resulting differential equations become almost identical to those of
the $R+aR^2+bR^3$ theory, the only differences being the values of the $\ld i$
and the factors $\al\Z3$ and $\be\Z3$, (which appear implicitly in (4.3) as the
powers involved in the roots).

We will not study the theory generated by the potential (4.12) in detail, but
remark that it should not lead to any differences from our former analysis any
more substantial than, say, the differences between the cubic and quadratic
theories. In particular, the uniqueness of the Schwarzschild solutions should
still apply to theories with $a>0$.\bigskip

We will now show that the uniqueness theorem does indeed apply to theories with
a polynomial $R$ action of arbitrary degree $n$, if $\ad2>0$. We note, first of
all, that the Einstein-scalar field equations with a general potential $\V$,
derived from the action (1.4a) are given by
$$\ze''=(m-1)^2\lbz-4\ka^2\V\gec,\eqno(4.13a)$$
$$\et''=m(m-1)\lbz-\left[{(m+1)4\ka^2\V\over m}+\ka\g1\dVds\right]\gec,
\eqno(4.13b)$$
$$\ch''=m(m-1)\lbz-\left[{(m+1)4\ka^2\V\over m}+\ka\g2\dVds\right]\gec,
\eqno(4.13c)$$
with the constraint
$$\eqalign{(m+1)\ze'^2+{2m\ze'(\g 2\et'-\g 1\ch')\over\gd}+{\mgB2\over(\gd)^2}
{\et'}^2-2{\mggB12\over(\gd)^2}\et'\ch'&\cr+{\mgB1\over(\gd)^2}{\ch'}^2+(m-1)
\lbz-{4\ka^2\V\over m}\gec&=0,\cr}\eqno(4.13d)$$
where we have used the coordinates (2.1), and functions $\ze$, $\et$ and $\ch$
defined by (2.3)--(2.5) with $\g1$ and $\g2$ as yet undetermined, (apart from
the assumption that $\g1\ne\g2$). In (4.13) $\si$ is assumed to be defined
implicitly by the inverse relation to (2.4) and (2.5), viz.
$$\ka\si={m\over2}\left(\ch-\et\over\gd\right).\eqno(4.14)$$
For a general polynomial $R$ action (1.1), (1.7) this implicit definition
becomes
$$\ee\left(2\ka\si\over\sqm\right)=\ee\exp\left(2(n-1)(\ch-\et)\over n\right)
=1+\spa p\ad pR^{p-1},\eqno(4.15)$$
on account of (1.2), while
$$\V={\ee\over4\ka^2}\exp\left(-4\g1\ka\si\over m\right)\spa(p-1)\ad pR^p,
\eqno(4.16)$$
and
$$\dVds={\ee\over m\ka}\exp\left(-4\g1\ka\si\over m\right)\spa\left(\g1-p\g3
\right)\ad pR^p,\eqno(4.17)$$
if we take $\g1$, $\g2$ and $\g3$ to be given by (2.10a) as before. Thus
if we define variables $X$, $Y$, $Z$, $V$ and $W$ by (2.7) we obtain the
5-dimensional autonomous system of first order differential equations
$$X'={-\ee Z^2\over m}\spa(p-1)\ad pR^p-{(m-1)P\over m}\,,\eqno(4.18a)$$
$$\eqalign{Y'&={-\ee Z^2\over m}\left\{\g1(\g1-\g3)R+\spa\left[\left(\GG-1
\right)-p\left(\g1\g3-1\right)\right]\ad pR^p\right\}-P\cr&={-\ee Z^2\over4(m+1
)}\left\{(m+2)R+\spa\left[m+2p\right]\ad pR^p\right\}-P,\cr}\eqno(4.18b)$$
$$\eqalign{V'&={-\ee Z^2\over m}\left\{\g2(\g1-\g3)R+\spa\left[\left(\g1\g2-1
\right)-p\left(\g2\g3-1\right)\right]\ad pR^p\right\}-P\cr&={-\ee Z^2\over4(k-1
)(m+1)}\left\{\left(2(k-1)-m\right)R+\spa\left[-(2k+m)+2(2k-1)p\right]\ad pR^p
\right\}-P,\cr}\eqno(4.18c)$$
$$Z'=YZ,\eqno(4.18d)$$
$$W'=VW,\eqno(4.18e)$$
where $R$ is now defined implicitly in terms of $W$ and $Z$ by
$$\ee\left(W\over Z\right)^{2(n-1)/n}=1+\spa p\ad pR^p,\eqno(4.18f)$$
and $P$ is still defined by (2.8f).

The properties of the phase space defined by (4.18) differ little in their
essentials from the examples studied earlier. We first note that an (anti)-de
Sitter subspace is defined by the (constant) values of $R$ given by roots of
the polynomial
$$m+\spa\left[m-2(p-1)\right]\ad pR^{p-1}=0.\eqno(4.19)$$
The 3-dimensional subspace is given by $V=Y$ and $W=\ga Z$, where by (4.18f)
$$\ee\ga^{2(n-1)/n}=1+\spa p\ad pR^p,\eqno(4.20)$$
$R$ being a solution of (4.19). The field equations of the subspace are once
again given by (2.15) with $\LA$ now defined by
$$\LA={\ee\over m}\spa(p-1)\ad pR^p.\eqno(4.21)$$
Our earlier results concerning the (anti)-de Sitter subspace follow through. In
particular, we are lead to the Schwarzschild-de Sitter type solutions (2.18),
which in terms of the original physical metric are given by
$$\dd s^2=-\DE\dd t^2+\DE^{-1}\dd r^2+r^2\gxxb,\eqno(4.22a)$$
with
$$\DE=\lb-{2GM\over r^{m-1}}-{\left[\spa(p-1)\ad pR^p\right]r^2\over m(m+1)
\left[1+\spa p\ad pR^p\right]}\,,\eqno(4.22b)$$
$R$ being a solution of (4.19). Furthermore, we once again obtain
Robinson-Bertotti solutions of the form (2.20) for our new definitions of $\ga$
and $\LA$.

Let us now consider the critical points of the system (4.18). It can be readily
seen from (4.18a-c)  that the only critical points at finite values of $X$,
$Y$, $V$, $Z$ and $W$ must all have: $P=0$, and also either (i) $W=Z=0$\foot
{$\;^{\hash9}$}{Ostensibly it would seem that we can simply take $Z=0$ here.
However, since $R$ is defined implicity in terms of
$W/Z$ the summation terms in (4.18a-c) will not vanish simply if the overall
factor of $Z^2$ vanishes. The leading order $R^n$ term within the summations is
of order $W^2Z^{-2}$, and so for consistency one must also require that $W=0$
when setting $Z=0$.}; or (ii) $R+\spa p\ad pR^p=0$ and $R+\spa\ad pR^p=0$. The
only value of $R$ which simultaneously satisfies both of the conditions (ii) is
$R=0$. We therefore retrieve the same critical points as were found in \S2 and
\S3, viz.\ \case (i): The familiar $W=0$, $Z=0$, $\lb=0$ surface discussed in
\S2. \case (ii): $X=Y=V=0$, $W=Z=Z_0$, ($R=0$). These are the points $\OZ$
which lie in the Schwarzschild subspace, discussed in \S3.

The phase space at infinity may be studied either by a direct analysis of the
4-sphere at infinity, using an approach similar to that described in \S2.2 and
\S2.3, or by the computationally more simple method of defining new variables
$\de$, $y$, $v$, $z$ and $w$ by
$$X={\pm1\over\de},\qquad Y={\pm y\over\de},\qquad V={\pm v\over\de},\qquad
Z={\pm z\over\de},\qquad W={\pm w\over\de},\eqno(4.23)$$
and classifying the $\de=0$ critical points of the resulting field equations
$$\pm\dta \de={\de\over m}\left[z^2f\Z1+(m-1)q\right],\eqno(4.24va)$$
$$\pm\dta y={1\over m}\left[(y-1)z^2f\Z1-\g1z^2f\Z2+q((m-1)y-m)\right],
\eqno(4.24b)$$
$$\pm\dta v={1\over m}\left[(v-1)z^2f\Z1-\g2z^2f\Z2+q((m-1)v-m)\right]
,\eqno(4.24c)$$
$$\pm\dta z={z\over m}\left[my+z^2f\Z1+(m-1)q\right],\eqno(4.24d)$$
$$\pm\dta w={w\over m}\left[mv+z^2f\Z1+(m-1)q\right],\eqno(4.24e)$$
where $\dd\ta=\de^{-1}\dd\xi$, $q\equiv\de^2P$ and
$$f\Z1\equiv\ee\spa(p-1)\ad pR^p,\eqno(4.24f)$$
$$f\Z2\equiv\ee\left[(\g1-\g3)R+\spa\left(\g1-p\g3\right)\ad pR^p\right],
\eqno(4.24g)$$
with $R$ now defined implicity in terms of $w$ and $z$. This method picks out
all critical points at infinity apart from those with $X=0$. To obtain all
critical points we must repeat the calculation with each of the phase space
coordinates in turn defined as $\pm1/\de$.

It is convenient to divide the critical points at infinity arising from this
analysis into three classes according to the behaviour of the scalar curvature
$R$ at the points: (i) $R=0$; (ii) $R=\const\ne0$; or (iii) $R\rarr\infty$.
\case (i): Critical points with $R=0$ have $f\Z1=0$ and $f\Z2=0$. Consequently
the only possibilities admitted by the equations (4.24) are: the points $L(y)$,
defined by (2.41), (2.42); and the points $M_{1,2}$, defined by (2.29) (with
$V=Y$ and $W=0$ also). If we repeat the above analysis for the system of
equations defined by putting $Z=1/\de$, with the other variables proportional
to $Z$, then for $R=0$ we also obtain the point $O_1$ defined by the $Z_0\rarr
\infty$ limit of (3.1). \case (ii): If $R=\const\ne0$ then we must have $z\ne0$
and $w\ne0$ on account of (4.18f). Equations (4.24) then imply that $v=y$ and
$f\Z2=0$. This latter condition is equivalent to (4.19) for $R\ne0$. Thus such
critical points must lie in the anti-de Sitter subspace, which as we have
already discussed is well-defined. By the analysis of \S2.2, these critical
points are therefore $S_{1,2}$ and $T_{1,2}$, where the $\ga$ and $\LA$ of the
defining relation (2.30b) and (2.31b) are now taken to be given by (4.20) and
(4.21) respectively. \case (iii):  Obviously points at which $R\rarr\infty$
cannot correspond to regular solutions. However, it is nevertheless interesting
to check whether the structure of the phase space is preserved when compared to
our earlier examples. If $R\rarr\infty$, then on account of (4.15)
$$R\goesas\pm\left\{{1\over n\ad n}\left[\ee\sig-1\right]\right\}^{1/(n-1)}.
\eqno(4.25)$$
In the limit $R\rarr\infty$ the field equations are therefore equivalent to
those derived from an exponential sum potential (1.10) consisting of a $n+1$
terms, with $\g1$, $\g2$ and $\g3$ given by (2.10a) and
$$\g j={2(n-1)+m(j-3)\over2(n-1)\sqm},\qquad 4\le j\le n+1.\eqno(4.26)$$
We could also derive the values of the appropriate coefficients, $\ld i$,
however, they are not of much concern to us here. We merely note that since
the field equations are equivalent to those of an exponential sum potential,
the only critical points at which $R\rarr\infty$ are precisely the points
$N_{1,2}$, $P_{1,2}$, $Q_{1,2}$ and $R_{1,2}$, (with appropriate $\ld i$ in the
definitions). The asymptotic form of the solutions is therefore given by
Table 1, or in terms of the physical metric, by Table 4. In particular, all
of these critical points correspond to asymptotic regions in which the
physical metric is not asymptotically flat.

Thus the only integral curves which join a critical point corresponding to an
horizon to a critical point corresponding to an asymptotically flat region
must be the trajectories with one endpoint on the curve formed by the
intersection of the $W=0$, $Z=0$, $\lb=0$ surface with the Schwarzschild
subspace, and with a second endpoint at $M_{1,2}$ or $O_1$. It therefore only
remains to determine the dimension of the set of such solutions. As before,
this may be found by a linearised analysis of small perturbations about the
points. This analysis will be unchanged from that of \S2 and \S3 since by
(4.15) we have to leading order in $R$
$$\ee\left(2\ka\si\over\sqm\right)=1+2aR+\OO(R^2).\eqno(4.27)$$
Consequently, the linearised perturbation equations will be identical to
those obtain in the \rsq. In particular, if $\ad2>0$ then no trajectories in
directions orthogonal to the Schwarzschild subspace will have endpoints at
$M_{1,2}$ or $O_1$. This completes our proof of the uniqueness theorem. We
have of course assumed that $\ad2\ne0$ throughout. If $\ad2=0$ then it seems
likely that the properties of the asymptotically flat solutions will be
essentially determined by the value of $\ad p$, where $p$ is the least value
such that $\ad p\ne0$.
\section 5. Discussion

To conclude, we have shown that in theories with an action polynomial in the
Ricci scalar, the only static spherically symmetric solution with a regular
horizon is the Schwarzschild solution, provided that the coefficient $\ad2$
of the quadratic term is positive. In fact, if we drop the condition of
regularity on the horizon then the only asymptotically flat solutions are still
the positive and negative mass Schwarzschild solutions. The generalised scalar
potential model of \S2 does possess non-Schwarzschild $W=0$, $Z=0$, $\lb>0$
solutions with naked singularities which approach flat space asymptotically
near the points $M_{1,2}$. However, these solutions correspond physically to
$\ld1=\ld2=\dots=\ld s=0$, a possibility which is not admitted in the higher
derivative theories. We have not considered the theories for which the
quadratic terms vanishes but which have non-zero terms at higher order. In such
cases the condition $\ad2>0$ presumably translates into a condition on the
coefficients of higher terms in the series. The question of solutions with
$\ad2<0$ is also not fully clear, and could perhaps be resolved by a numerical
study.

The fact that we have been able to prove a uniqueness theorem for static
spherically symmetric black holes in the case that $\ad2>0$ has immediate
important physical implications when considered in conjunction with the earlier
work of Pechlaner and Sexl [19] and Michel [20]. These authors observed
that in the case of the \rsqfd\ the solutions with $R=0$ are {\it not}
the solutions which correspond to the weak field limit about any physical body
such as a star or point particle. This is also the case for the general theory
with a polynomial $R$ action which we are considering here. Suppose, for
example, that we wish to match our solutions onto a star with energy-momentum
tensor $T_{ab}$ in its interior. If we add such a term to (1.8) and trace the
result we find that
$$\eqalign{{-1\over2}(&D-2)R+\spa\ad p\left\{\left(p-{D\over2}\right)R^p+(D-1)
p(p-1)R^{p-3}\Bigl(R\Dal R+(p-2)R^{;c}R_{;c}\Bigr)\right\}\cr&=2\ka^2T.\cr}
\eqno(5.1)$$
Thus if we set $T=0$ at the surface of the star we cannot conclude, as in the
case of the Schwarzschild solution, that $R=0$. Our results show, however, that
if $\ad2>0$ and $R\ne0$ anywhere in the \doc, then the solutions are {\it not
asymptotically flat}.

Pechlaner and Sexl, on the contrary, assumed the existence of
non-Schwarzschild solutions which are asymptotically flat. Our full non-linear
analysis seems to invalidate this assumption in the $\ad2>0$ case.
Consequently, their weak field analysis of the \rsqfd, in which they derived
experimental bounds on the parameter $a$ from resulting fifth-force type
effects, needs to be reexamined with appropriately changed boundary conditions.

If the Schwarzschild solutions are not the solutions of physical interest in
these models, it would seem that the large class of asymptotically
anti-de Sitter and de Sitter solutions hold more promise. These include both
the exact solutions (2.18), (with $\LA$ given, for example, by (3.12) and
(4.11) in the quadratic and cubic order theories respectively), and also
other solutions asymptotic to them at infinity, or at the de Sitter
cosmological horizon. Asymptotically (anti)-de Sitter solutions have also been
found in a number of models in $D>4$ dimensions which incorporate a
Gauss-Bonnet term [32-34] and other dimensionally continued Euler densities
[35], and thus appear to be a generic feature of higher derivative theories. In
fact, maximally symmetric solutions have been found to exist in a much wider
class of higher derivative models [31], so presumably such models also possess
asymptotically (anti)-de Sitter black hole solutions.

Returning to the models with dimensionally continued Euler densities, we note
that similarly to the solutions found here, only asymptotically anti-de Sitter
solutions are found if the coefficient of the Gauss-Bonnet term is positive in
the quadratic theory, while a de Sitter branch can be obtained at higher order
[35].  One important difference between our solutions and those of Boulware and
Deser [32], for example, is that the asymptotically anti-de Sitter branch of
the Boulware-Deser solutions has a negative gravitational mass, giving rise
to an instability. Our solutions have a positive gravitational mass, and so the
question of their stability is still an open problem. The issue of the weak
field limit has not been addressed in refs.\ [32-34], since of course it is
really only relevant in compactified models.

If the de Sitter sector is to be treated as a serious physical model it is
clear that the effective cosmological term must be small to be consistent with
observation. It is interesting to note that recent astronomical evidence
actually favours a small positive cosmological constant\foot{$^{\hash10}$}{The
``best-fit'' value favoured in ref.\ [36] is $\LA=3.1\times10^{-52}m^{-2}$, or
$\LA=8.1\times10^{-122}$ in dimensionless Planck units.} [36]. If higher order
terms in $R$ are obtained from the dimensional reduction of an action
corresponding to the low energy limit of a higher-dimensional string theory,
then the status of the effective cosmological term is at best uncertain. If we
ignore the values of the dilaton and compacton and assume that the
compactification scale is approximately of order $\sqrt{\al'}$ (in units in
which $c=\hbar=1$, as used throughout this paper), where $\al'$ is the Regge
slope parameter, then the coefficients $\ad p$ are of order $(\elp)^{2(p-1)}$,
and $\LA$ is of order $(\elp)^{-2}$, to within a few orders of magnitude. Such
a colossally large effective cosmological term would of course spell disaster
for these models. However, no definite statements can be made without some
knowledge of the expectation values of the dilaton and compacton fields, both
of which couple non-trivially to the higher order curvature terms in four
dimensions. In fact, these scalar fields should really be treated dynamically,
which would necessitate a complete reexamination of the model. In the
$D>4$ uncompactified quadratic order theory with a Gauss-Bonnet term the
dilaton has the effect of removing the anti-de Sitter branch [34], but it is
not clear how the dilaton would affect compactified models.

In view of these problems it would also be interesting, especially in the
quadratic order case in four dimensions, to consider the addition to the
lagrangian of a term comprising the square of the Ricci tensor. In that case
the effective theory contains a massive spin 2 field with a non-trivial
coupling to gravity in addition to the scalar field. The effective energy
momentum tensor of the extra excitations does not satisfy criteria usually
required in the proof of the no-hair theorems, and thus it seems plausible that
the no-hair theorems could be circumvented in such a model. However, the
dynamical system arising from a static spherically symmetric ansatz for the
metric and the other fields is considerably more complicated than in the models
we have studied in this paper, and it is not clear to us whether it can be
reduced to a tractable form.

We remark parenthetically that the results of this paper seem to hold up some
hope for the problem of finding black hole solutions in dimensionally reduced
theories in which the internal space is non-Ricci flat, since such models can
also be treated by the formalism of \S2 and \S3. In I and II
the potential corresponding to (1.10) was limited to two terms at most.
However, we have seen that at least three exponential terms are required in the
potential in order to obtain a Schwarzschild subspace, and thus no
asymptotically flat solutions were found in I and II for internal spaces of
non-zero curvature. More complicated models with $s\ge3$, (or some equivalent
condition if more than one scalar field is present), may yield more interesting
results.
\section Appendix A

In the case of the quadratic action $R+aR^2$ in four dimensions with $a>0$,
Whitt [21] has shown that provided the energy momentum of any additional
matter fields satisfies certain conditions then the usual no hair theorem for
stationary, axisymmetric, asymptotically flat black holes [37] is still
valid. He was able to obtain this result by proving that the curvature scalar
for such solutions vanishes in the \doc, and therefore the effective theory
coincides with the usual Einstein-Hilbert theory. Unfortunately his proof
cannot be generalised to more general polynomial actions of the form (1.7). For
completeness we will present an account of his proof here, in order to show
why the theorem cannot be directly extended to more general polynomial $R$
actions, and also to correct some errors present in the original paper.

Whitt's proof is based on a study of the 4-dimensional Killing bivector [38]:
$$\rh_{ab}=2m_{[a}k_{b]},\eqno(A.1)$$
where $m^a$ is the spacelike Killing vector with period $2\pi$ and $k^a$ is
the timelike Killing vector normalized to unity at infinity. These vector
fields commute [39]:
$$k^b m^a_{\ ;b}-m^b k^a_{\ ;b}=0.\eqno(A.2)$$
We will take $l^a$ to denote either of the Killing vectors $k^a$ and $m^a$.
Making use of (A.2) and the identity
$$(l_{[a;b}l_{c]})^{;c}={2\over 3}l^c R_{c[a}l_{b]}\,,\eqno(A.3)$$
which is valid for any Killing vector [40], we find
$$\lkm =-{1\over 2}l^dR_{d[a}\kab.\eqno(A.4)$$
We can now use the vacuum field equations of the higher derivative theory to
evaluate the right hand side of this expression. For the $R+aR^2$ case one
finds
$$\lkm={-a\,l^dR_{;d[a}\kab\over 1+2aR},\eqno(A.5a)$$
or in the general case, with field equations (1.8),
$$\lkm={-\left(\spa\ad pp\,l^dR_{;d[a}\kab\right)\over2\left(1+\spa\sap\right)}
.\eqno(A.5b)$$
Using the orthogonality of the Killing vectors to $R_{;a}$ and the
antisymmetry of the derivatives of the Killing vectors, these expressions
become respectively
$$\lkm={-aR^{;d}l_{[d;a}\kab\over1+2aR}\eqno(A.6a)$$
and
$$\lkm={-\spa\psa^{;d}l_{[d;a}\kab\over2\left(1+\spa\sap\right)}\eqno(A.6b)$$
The solution of equations (A.6) is given respectively by
$$l_{[a;b}k_c m_{d]} =C(1+2aR)^{-1/2}\ep_{abcd}\,\eqno(A.7a)$$
and
$$l_{[a;b}k_c m_{d]}=C\left(1+\spa\sap\right)^{-1/2}\ep_{abcd}\,\eqno(A.7b)$$
where $C$ is an arbitrary constant. Now we know that the left hand side of
(A.7) vanishes on the axis of rotation since $m^a=0$ there. On account of the
asymptotic flatness of the solutions $R$ cannot be singular everywhere on the
axis, and hence C must vanish giving
$$k_{[a;b}\rh_{cd]}=0,\qquad\qquad m_{[a;b}\rh_{cd]}=0,\eqno(A.8)$$
throughout the \doc.

{}From (A.8) it follows [40] that if the \doc\ is simply connected and admits
no closed timelike curves, then $\rh_{ab}$ is timelike throughout the \doc,
becoming null on its inner boundary, which is a null hypersurface.

At this point Whitt [21] considers the trace of the field equations in the
absence of matter, which for the quadratic theory yield
$$-6aR^{;a}_{\ \;;a}+R=0.\eqno(A.9)$$
One can now multiply by $R$ and integrate over the \doc\ to obtain
$$6a\int R\Rca d\Sigma^a = 6a\int R^{;a}\Rca\gdx+\int R^2\gdx\,.\eqno(A.10)$$
The left hand side vanishes because $\Rca$ is orthogonal to the inner
boundary and is zero on the outer boundary due to asymptotic flatness.
Moreover the right hand side is positive definite, provided $a>0$, since $\Rca$
cannot be timelike anywhere in the \doc\ by virtue of being orthogonal to $\rh_
{ab}$. Hence $R$ must be identically zero there.

The conformally rescaled theory coincides therefore with the usual Einstein
theory since the scalar field $\sigma$ is clearly zero if $R$ vanishes and it
follows that the uniqueness theorem for stationary, axisymmetric,
asymptotically flat solutions still holds. Moreover, it is easy to see that
the result remains valid when additional matter fields are included in the
action, provided that the energy momentum tensor is traceless, $T^a_{\ a}=0$,
and satisfies the matter circularity condition [40] $T_{d[a}\kab=0$. In such
cases the usual no-hair theorems are still valid. These conditions are always
satisfied, in particular, by a stationary, axisymmetric electromagnetic field.
Unfortunately, however, Whitt's result cannot be extended to the general
polynomial $R$ action since in that case the argument for $R$ being zero in the
\doc\ breaks down. Specifically, the trace of the field equation now becomes
$$-R+\spa\ad p\left[(p-2)R^p+3p(R^{p-1})^{;a}_{\ \;;a}\right]=0\eqno(A.11)$$
and therefore, integrating by parts and discarding the boundary term as before
we find
$$-\int R^2\gdx +\spa (p-2) \ad p\int R^{p+1}\gdx =\spa 3p(p-1)\ad p\int R^{p-2
}\Rca R^{;a}\gdx\eqno(A.12)$$
The integrals in the sum on the left hand side do not have a definite sign for
even $p$, while the integrals in the sum on the right hand side do not have a
definite sign for odd $p$. Thus it is impossible to conclude from this relation
that $R$ must vanish in the general case, or indeed in any special case other
than $n=2$, $\ad2>0$.
\section Appendix B

We list here the exact solutions obtained in the cases (i) $\V\equiv0$, i.e.,
$\ld1=\ld2=\dots\ld s=0$ -- appropriate to the $W=0$, $Z=0$ subspace; (ii)
$s=1$, $\lb=0$ -- appropriate to the $\lb=0$ surface in the $W=0$ subspace.
Corresponding solutions for the $\lb=0$ surface in the $Z=0$ subspace may be
obtained by substituting $\g1\rarr\g2$, $\ld1\rarr\ld2$.
\subsection B.1 Solutions with $W=0$ and $Z=0$

The one-parameter family of solutions was derived in I. In that case solutions
were parametrised in terms of the number $n_e$ of extra dimensions in the \KK\
model. Here a parametrisation in terms of $\g1$ (or $\g2$) is more natural. We
will use the former parametrisation. We find
$$e^\ze={CA_1\over\Dd1}\expm\,,\eqno(B.1)$$
$$\et={m\over m-1}(\ze+k\xi)+\const\eqno(B.2)$$
$$\eu=A_0e^{-\ad0\xi},\eqno(B.3)$$
$$\ev={(m-1)CA_1\expm\over(m-1)CA_1^{\ 2}\exp\left[(m-1)C\ze\right]+\left(\ad0+
\mC\right)\Dd1}\,,\eqno(B.4)$$
$$\rH^{m-1}={CA_1\over A_0\Dd1}\exp\left[\left(\ad0+{1\over2}(m-1)C\right)\xi
\right],\eqno(B.5)$$
$${2\g1\ka\si\over m}={(m-1)\g1^{\ 2}(\cd1-mk)\over\mg1}+\const,\eqno(B.6)$$
where $$\Dd1\equiv\lb-A_1\Y{\ 2}\exp\left[(m-1)C\ze\right]\,,\eqno(B.7)$$
while $k$, $A_0$ and $A_1$ are arbitrary constants, $C$ is a non-zero constant
given by
$${1\over4}(m-1)^2\mgB1C^2=m^2k^2+(m-1)\cd1^{\ 2}\g1^{\ 2},\eqno(B.8)$$
and the constant $\ad0$ is defined by
$$\ad0={mk+(m-1)\g1^{\ 2}\cd1\over\mg1}\,.\eqno(B.9)$$
and lies in the range $-{1\over2}(m-1)|C|\le\ad0\le{1\over2}(m-1)|C|$.
Solutions have an asymptotic ($\rH\rarr\infty$) region only if $\lb>0$.

As $\xi\rarr-\infty$ we find $\rH\rarr0$, giving rise to a singularity
except in the special case $\cd1=mk=\ad0=-{1\over2}(m-1)|C|$, when $\rH\rarr
\const$, suggesting the possible presence of an horizon. This can be
immediately verified to be true since (B.5) can be inverted. If we choose\foot
{$^{\hash11}$}{Equations (B.10)-(B.12) correct a sign discrepancy in I for the
$\lb<0$ and $\xi\rarr+\infty$ cases.}
$$A_1=\cases{\dsp{\lb A_0/|\lb|},&$C>0$,\cr\dsp{|\lb|/A_0},&$C<0$,\cr}
\eqno(B.10)$$
we find
$$\euu=e^{-2\vv}=\lb\left(1-{|C|\ \over|\lb|\rH^{m-1}}\right),\eqno(B.11)$$
which correspond to the domain of outer communications of the positive mass
Schwarzschild solution for $\lb>0$. Similarly, as $\xi\rarr+\infty$ we find
$\rH\rarr0$, giving rise to a singularity, unless $\cd1=mk=\ad0={1\over2}(m-1)
|C|$. In that case we once again obtain (B.11) if we choose
$$A_1=\cases{\dsp{-|\lb|/A_0},&$C>0$,\cr\dsp{-\lb A_0/|\lb|},&$C<0$,
\cr}\eqno(B.12)$$
\subsection B.2 Solutions with $W=0$ and $\lb=0$

These solutions may be derived as in I. If $\GG<m+1$ we find
$$e^\et={\CB B_1\over\Dd2}\expc\,,\eqno(B.13)$$
$$e^\ze=B_0\exp\left[{m(\et+\kb\xi)\over\Mg1}\right]\,,\eqno(B.14)$$
$$\eu=B_0\Y{\ m}B_2^{\ (m-1)}\left\{{\CB B_1\over\Dd2}\exp\left[\left(\Cb+m\kb-
(m-1)\bd0\right)\xi\right]\right\}^{1/(\Mg1)},\eqno(B.15)$$
$$\ev={\MgB1B_0\left\{\Dd2^{1-\GG}\left(\CB B_1\exp\left[\left(\Cb+\kb\right)
\xi\right]\right)^m\right\}^{1/(\Mg1)}\over\CB B_1^{\ 2}\exp\left[\CB\xi\right]
+\left(\Cb+\bd0\right)\Dd2}\,,\eqno(B.16)$$
$$\rH={1\over B_0B_2}\left\{{\CB B_1\over\Dd2}\exp\left[\left(\Cb+\bd0\right)
\xi\right]\right\}^{1/(\Mg1)},\eqno(B.17)$$
$$\exp\left(2\g1\ka\si\over m\right)={1\over B_2}\left\{{\CB B_1\over\Dd2}\exp
\left[\left(\Cb+{1\over\GG}(m\kb+\bd0)\right)\xi\right]\right\}^{\GG/(\Mg1)},
\eqno(B.18)$$
where $$\Dd2\equiv\LEB-B_1\Y{\ 2}\exp\left[\CB\xi\right]\,,\eqno(B.19a)$$
with $$\LEB\equiv{1\over m}\MgB1\ld1,\eqno(B.19b)$$
while $B_0$, $B_1$, $B_2$ and $\kb$ are arbitrary constants, $\CB$ is a
non-zero constant given by
$${1\over4}\mgB1\CB^2=m^2\kb^2+\MgB1\GG\cd1^{\ 2},\eqno(B.20)$$
and the constant $\bd0$ is defined by
$$\bd0={(\GG-1)m\kb+\MgB1\GG\cd1\over \mg1}\,.\eqno(B.21)$$
and lies in the range $-\HCBM\le\bd0\le\HCBM$.

We now find that the limit $\xi\rarr-\infty$ corresponds to $\rH\rarr0$ except
in the special instances when $m\kb/(\GG-1)=\cd1=\bd0=-\HCBM$, for which
$\rH\rarr\const$, suggesting the possible presence of an horizon. This indeed
the case: (B.17) can be inverted and if we make the choice
$$B_0^{\ \GG}B_2^{\ \GG-1}={1\over\Mg1},\eqno(B.22)$$
$$B_1=\cases{\dsp B_0^{\ m+1}B_2^{\ m},&$\CB>0$,\cr\dsp{\LEB/(B_0^{\ m+1}B_2^
{\ m})},&$\CB<0$,\cr},\eqno(B.23)$$
then we find the solution
$$\dd s^2=-\rH^2\DEB\dd t^2+\rH^{2(\GG-1)}{\dd\rH^2\over\DEB}+\rH^2\gxxb,\eqno
(B.24a)$$
where
$$\DEB=\LEB\left(1-{|\CB|\qquad\qquad\over|\LEB|\MgB1\rH^{\Mg1}}\right)\,,\eqno
(B.24b)$$
while the scalar field is given by
$$\exp\left(2\g1\ka\si\over m\right)={\rH^{\GG}\over\Mg1}\,.\eqno(B.24c)$$
Similarly the limit $\xi\rarr+\infty$ also corresponds to $\rH\rarr0$ except
for the special cases when $m\kb/(\GG-1)=\cd1=\bd0=\HCBM$
which $\rH\rarr\const$ Equation (B.17) can once again be inverted, and we
retrieve the solution (2.18) if we now make the choice (B.22) and\foot{$^{\hash
12}$}{There are some sign discrepancies in I and II for the expressions
corresponding to (B.23), (B.24b) and (B.25).}
$$B_1=\cases{\dsp{-\LEB/(B_0^{\ m+1}B_2^{\ m})},&$\CB>0$,\cr\dsp-B_0^{\ m+1}B_
2^{\ m},&$\CB<0$,\cr},\eqno(B.25)$$
The spacetimes thus have naked singularities except in the special cases above.

An asymptotic region is defined only for $\ld1>0$: $\rH\rarr\infty$ when $\CB
\xi=\ln|\LEB/B_1^2|$. This limit is approached at the points $N_{1,2}$ at
infinity (c.f.\ (2.37)). All $\ld1>0$ solutions have one endpoint at $N_1$ or
$N_2$, and another endpoint on the $\ld1=0$ curve of critical points. The
$\ld1<0$ solutions have one endpoint on the $\ld1=0$ curve in the first
quadrant, and another endpoint on the $\ld1=0$ curve in the third quadrant.
The asymptotic form of all the $\ld1>0$ solutions (B.15)-(B.18) is given by
$$\euu\goesas\rH^2,\qquad \evv\goesas\rH^{2(\GG-1)},\qquad e^{2\ka\si}\goesas
\rH^{m\g1}.\eqno(B.26)$$
Thus none of the solutions is asymptotically flat. The general solution holds
for $\CB\ne0$. If one sets $\CB=0$ while integrating the differential equations
one finds a solution which corresponds to the $\CB=0$ limit of (B.24).\medskip

If $\GG>m+1$ then the solutions given by expressions (B.13)-(B.21) are still
valid but $\bd0$ now lies in the range $\bd0\le-\HCBM$, $\bd0\ge\HCBM$.
Furthermore, the behaviour of the solutions and the direction of trajectories
near the critical points is greatly changed in some instances.
We now find that the limit $\xi\rarr-\infty$ corresponds to $\rH\rarr0$ as
before if $\bd0<-\HCBM$. However, if $\bd0\ge\HCBM$ we find that $\rH\rarr
\infty$. Similarly, the limit $\xi\rarr+\infty$ corresponds to $\rH\rarr0$ if
$\bd0>\HCBM$, and $\rH\rarr\infty$ if $\bd0\le-\HCBM$. Critical points in the
first and third quadrants have $\rH\rarr0$, while those in the fourth and
second quadrants have $\rH\rarr\infty$. Since each solution has a different
endpoint in the $\rH\rarr\infty$ regime, the asymptotic behaviours vary. We
find
$$\euu\goesas\rH^{2\uu\Z0},\qquad\evv\goesas\rH^{2\vv\Z0},\qquad\exp\left(2\g1
\ka\si\over m\right)\goesas\rH^{2\si\Z0},\eqno(B.27a)$$
where
$$\uu\Z0={\pm\HCBM+m\kb-(m-1)\bd0\over\bd0\pm\HCBM},\eqno(B.27b)$$
$$\vv\Z0={m\left(\kb\pm\HCBM\right)\over\bd0\pm\HCBM},\eqno(B.27c)$$
$$\si\Z0={\pm\HCBM+{1\over\GG}\left(m\kb+\bd0\right)\over\bd0\pm\HCBM},\eqno
(B.27d)$$
and the upper (lower) sign refers to $\xi\rarr-\infty$ ($\xi\rarr+\infty$).
For the limiting case solutions with $\bd0={1\over2}|\CB|$, if $\xi\rarr-\infty
$, or $\bd0=-{1\over2}|\CB|$, if $\xi\rarr+\infty$ we obtain the asymptotic
behaviour appropriate to (B.24), viz.
$$\euu\goesas\rH^{\,\GG-m+1},\qquad \evv\goesas\rH^{\,\GG+m-1},\qquad e^{2\ka
\si}\goesas\rH^{\,m\g1}.\eqno(B.28)$$
The limit, $\CB\xi=\ln|\LEB/B_1^{\ 2}|$ for the $\ld1<0$ solutions, which is
reached at the points $N_{1,2}$, now corresponds to $\rH\rarr0$. The $\ld1<0$
solutions given by (B.13)-(B.21) all have one endpoint at $N_1$ or $N_2$ and
another endpoint on the $\ld1=0$ curve. If this second endpoint is in the
fourth or second quadrant the solutions have an asymptotic region, otherwise
they do not. The $\ld1>0$ solutions, on the other hand, all have asymptotic
regions: they now describe trajectories with one endpoint on the $\lb=0$ curve
in the first (or third) quadrant, and a second endpoint on the same curve in
the fourth (or second) quadrant.

In addition to the above solutions, a second group of solutions also exist if
$\GG>m+1$ and $\ld1<0$. These solutions are given by
$$e^\et={|\CT|\over\LEB^{\ 1/2}\cosxi}\,,\eqno(B.29)$$
$$e^\ze=B_0\exp\left[{-m(\et+\kb\xi)\over\mG1}\right]\,,\eqno(B.30)$$
$$\eu=B_0\Y{\ m}B_2^{\ (m-1)}\left\{\lccos\exp\left[\left((m-1)\bd0-m\kb\right)
\xi\right]\right\}^{1/(\mG1)},\eqno(B.31)$$
$$\ev={-\left(\mG1\right)B_0\over\bd0+\CT\tan\left(\CT(\xi-\xi\Z0\right)}\left
\{\lccos\exp(-\kb\xi)\right\}^{m/(\mG1)}\,,\eqno(B.32)$$
$$\rH={1\over B_0B_2}\left\{\lccos\exp(-\bd0\xi)\right\}^{1/(\mG1)},\eqno(B.33)
$$
$$\exp\left(2\g1\ka\si\over m\right)={1\over B_2}\left\{\left[\lccos\right]^{
\GG}\exp\left[-(m\kb+\bd0)\xi\right]\right\}^{1/(\mG1)},\eqno(B.34)$$
where $\LEB$ and $\bd0$ are defined as before, $\xi\Z0$, $B_0$, $B_2$ and $\kb$
are arbitrary constants, and $\CT$ is a
non-zero constant given by
$$\mgB1\CT^2=\left(\mG1\right)\GG\cd1^{\ 2}-m^2\kb^2.\eqno(B.35)$$
These solutions have no asymptotic region, and correspond to trajectories
with one endpoint at $N_1$ and a second endpoint at $N_2$. These endpoints are
reached when $\xi=\xi\Z0+(2n+1)/(2\pi|\CT|)$, for integer $n$.\medskip

If $\GG=m+1$ we find the solutions
$$e^\et=B_1\CH^2e^{\CH\xi/2}\,,\eqno(B.36)$$
$$e^\ze=B_0\exp\left[\kb\xi+\ld1B_1e^{\CH\xi}\right]\,,\eqno(B.37)$$
$$\eu=B_0\Y{\ m}B_2^{\ (m-1)}\left\{\exp\left[\left(\kb-(m-1)\bd0\right)\xi
+\ld1B_1e^{\CH\xi}\right]\right\}^{1/m},\eqno(B.38)$$
$$\ev={mB_0\exp\left[\kb\xi+\ld1B_1e^{\CH\xi}\right]\over\kb+\bd0+\ld1B_1\CH
e^{\CH\xi}}\,,\eqno(B.39)$$
$$\rH={1\over B_0B_2}\left\{\exp\left[(\kb+\bd0)\xi+\ld1B_1e^{\CH\xi}\right]
\right\}^{1/m},\eqno(B.40)
$$
$$\exp\left(2\g1\ka\si\over m\right)={1\over B_2}\left\{\exp\left[\left(\kb+{1
\over m+1}(\bd0-{m\CH\over2})\right)\xi+\ld1B_1e^{\CH\xi}\right]\right\}^{(m+1)
/m},\eqno(B.41)$$
where $B_0$, $B_1$ $B_2$ are arbitrary constants,
$$\bd0={1\over m}\left({\CH\over2}+(m+1)\cd1\right),\eqno(B.42)$$
$\kb$ is a constant which lies in the range $|\kb|>\sqrt{m^2-1}\,|\cd1|/m$,
and $\CH$ is a non-zero constant given by
$$\CH=\left(2\over m-1\right)\left[m\kb\pm\sqrt{m^2\kb^2-(m^2-1)\cd1^{\ 2}}\,
\right],\eqno(B.43)$$
and has the same sign as $\kb$. We now find that for $\kb>0$ as $\xi\rarr-
\infty$ $\rH\rarr0$ except in the case that $\kb=-\cd1=-\bd0=\CH/2$, when
$\rH\rarr\const$ The same is true for $\kb<0$ in the limit $\xi\rarr+\infty$.
On the other hand, if $\kb<0$ and $\xi\rarr-\infty$, or if $\kb>0$ and
$\xi\rarr+\infty$, then $\rH\rarr\infty$. All solutions have an asymptotic
region. In terms of the phase space, all trajectories approach the points
$N_{1,2}$ at infinity which coincide with points $L_{5,7}$. In the case that
$\kb=-\cd1=-\bd0=\CH/2$ we can invert (B.40), and if we make the
choice
$$B_2=B_0^{\ -1},\qquad B_1={mB_0^{\ 2}\over\ld1\CH}\,,\eqno(B.44)$$
we obtain the solution
$$\dd s^2=-\rH^2\CH\ln\rH\dd t^2+{\rH^{2m}\dd\rH^2\over\CH\ln\rH}+\rH^2\gxxb,
\eqno(B.45a)$$
with scalar field
$$\exp\left(2\g1\ka\si\over m\right)=B_0\rH^{m+1}.\eqno(B.45b)$$
\bigbreak\leftline{\bf Acknowledgements\message{Acknowledgements}}\nobreak
\smallskip\noindent SM would like to thank M. Cadoni and M. Ferraris for
helpful discussions during the early stages of work on the problem, and the
Accademia dei Lincei for financial support. DLW would like to thank J. Barrow,
A. Burd and J. McCarthy for discussions and the Rothmans Foundation for
financial support.

\vfil\eject\leftline{\bf Figure captions\message{Figure captions}}\nobreak
\smallskip\noindent {\bf Fig.\ 1} Trajectories in the $W=0$, $Z=0$, $V=Y$
plane. The bold lines $Y=X$ and $Y=(m+1)X/(m-1)$ represent critical points
which respectively correspond to the horizons of the positive mass
Schwarzschild solutions (dashed lines), and the singularities of the negative
mass Schwarzschild solutions. The trajectory through the origin is flat space.

\noindent {\bf Fig.\ 2} The plane $V=Y=X$, $W=\ga Z$, with radial coordinate
$\rhb$ and angular coordinate $\th$: {\bf(a)} $\LA<0$, $\lb<0$; {\bf (b)}
$\LA>0$, $\lb>0$. The solutions are given in (2.20) in terms of $Z=\rhb\ct/(
1-\rhb)$.

\noindent {\bf Fig.\ 3} The hemisphere at infinity for the 3-dimensional
(anti)-de Sitter subspace: {\bf (a)} $\LA<0$; {\bf (b)} $\LA>0$.

\noindent {\bf Fig.\ 4} The projection of trajectories in the $W=0$, $Z=0$
subspace onto the $X$,$Y$ plane, with $V$ given by (2.12), for non-zero $\cd1$:
{\bf (a)} $\GG<m+1$; {\bf (b)} $\GG>m+1$. The broken line corresponds to the
Schwarzschild solution. The bold lines represent sections through the cone of
critical points $\lb=0$, $W=0$, $Z=0$. If $\cd1=0$ we obtain the section which
bisects the cone, so that instead of being hyperbolae the critical points fall
on the lines $\dsp Y={1\over m-1}\left[m\pm{\mgB1}^{1/2}\right]$.

\noindent {\bf Fig.\ 5} The projection of trajectories in the $W=0$, $\lb=0$
subspace onto the $X$,$Y$ plane, with $V$ given by (2.12), for non-zero $\cd1$:
{\bf (a)} $\GG<m+1$; {\bf (b)} $\GG=m+1$; {\bf (c)} $\GG>m+1$. The bold lines
represent the same critical points as in Fig.~4.

\noindent {\bf Fig.\ 6} The hemisphere at infinity for the 3-dimensional
$W=0$ subspace with $\ld1>0$: {\bf (a)} $\GG<1$; {\bf (b)} $\GG=1$; {\bf (c)
} $1<\GG<m+1$; {\bf (d)} $\GG\ge m+1$.

\noindent {\bf Fig.\ 7} The hemisphere at infinity for the 3-dimensional
$W=0$ subspace with $\ld1<0$: {\bf (a)} $\GG\le m+1$; {\bf (b)} $\GG>m+1$.

\noindent {\bf Fig.\ 8} The plane $Y=X$ or $Y=(m+1)X/(m-1)$ in the
Schwarzschild subspace. Both axes are 1-parameter families of critical points.
The solution in the $Y=X$ plane  corresponds to the $\lb=0$, $\LA=0$ limit of
(2.20), while the solution in the $Y=(m+1)X/(m-1)$ plane corresponds to the
limit $\g1=0$, $\LA_1=0$ of the solutions (B.13)-(B.21) in Appendix B.

\noindent {\bf Fig.\ 9} The planes $Y=m(X+k)/(m-1)$ in the Schwarzschild
subspace: {\bf (a)} $k=0$; {\bf (b)} $k<0$. The dashed trajectories are
positive mass Schwarzschild solutions. If $k>0$ these trajectories are
located in the $X<-m|k|$ region.

\noindent {\bf Fig.\ 10} The hemisphere at infinity for the 3-dimensional
Schwarzschild subspace: {\bf (a)} with $\th$, $\ph$ defined by (2.24); {\bf (b)
} with $\th$, $\ph$ defined by (3.4).
\references{[1] H. Weyl, \AP{Leipzig}{59} (1919) 101; Phys.\ Zeit.\ {\bf 22}
(1921) 473.\cr [2] K.S. Stelle, \PR{D16} (1977) 953; \GRG{9} (1977) 353.\cr
[3] I. Antoniadis and E.T. Tomboulis, \PR{D33} (1986) 2756.\cr
[4] D.A. Johnston, \NP{B297} (1988) 721.\cr
[5] A.A. Starobinsky, \PL{91B} (1980) 99.\cr
[6] R. Kerner, \GRG{14} (1982) 453.\cr
[7] A.A. Starobinsky, Sov.\ Astron.\ Lett.\ {\bf9} (1983) 302.\cr
[8] J.D. Barrow and A.C. Ottewill, \JP{A16} (1983) 2757.\cr
[9] M.B Miji\'c, M.S. Morris and W.-M. Suen, \PR{D34} (1986) 2934;\br A.A.
Starobinsky and H.-J. Schmidt, \CQG{4} (1987) 695;\br A.L. Berkin, \PR{D42}
(1990) 1016.\cr
[10] S.W. Hawking and J.C. Luttrell, \NP{B247} (1984) 250.\cr
[11] G.T. Horowitz, \PR{D31} (1985) 1169.\cr
[12] H.F. Dowker and R. Laflamme, \NP{B366} (1991) 209.\cr
[13] B. Zwiebach, \PL{156B} (1985) 315;\br B. Zumino, Phys. Rep. {\bf 137}
(1986) 109.\cr
[14] J. Madore, \PL{110A} (1985) 289;\br F. M\"uller-Hoissen, \PL{163B} (1985)
106;\br S. Mignemi, \MPL{A1} (1986) 337.\cr
[15] R.C. Myers, \NP{B289} (1987) 701.\cr
[16] A.L. Berkin and K. Maeda, \PL{245B} (1990) 348.\cr
[17] Q. Shafi and C. Wetterich, \PL{129B} (1983) 387; {\bf 152B} (1985) 51\cr
[18] K. Maeda, \PR{D37} (1988) 858; \PR{D39} (1989) 3159;\br
J.D. Barrow and S. Cotsakis, \PL{258B} (1991) 299.\cr
[19] E. Pechlaner and R. Sexl, \CMP{2} (1966) 165.\cr
[20] F.C. Michel, \AP{N.Y.}{76} (1973) 281.\cr
[21] B. Whitt, \PL{145B} (1984) 176.\cr
[22] P.W. Higgs, \NC{11} (1959) 816.\cr
[23] G. Magnano, M. Ferraris and M. Francaviglia, \GRG{19} (1987) 465;
\CQG{7} (1990) 557;\br A. Jakubiec and J. Kijowski, \PR{D37} (1988) 1406;\br
J.D. Barrow and S. Cotsakis, \PL{214B} (1988) 515.\cr
[24] B. Shahid-Saless, \PR{D35} (1987) 467.\cr
[25] H. Buchdahl, \NC{23} (1961) 91.\cr
[26] J.E. Chase, \CMP{19} (1970) 276;\br J.D. Bekenstein, \PR{D5} (1972) 1239;
{\bf D5} (1972) 2403.\cr
[27] G.W. Gibbons, ``Self-gravitating magnetic monopoles, global monopoles and
black holes'', in J.D. Barrow, A.B. Henriques, M.T.V.T. Lago and M.S. Longair
(eds.), {\it The Physical Universe: Proceedings of the XII Autumn School,
Lisbon, 1990}, (Springer, Berlin, 1991).\cr
[28] S. Mignemi and D.L. Wiltshire, \CQG{6} (1989) 987.\cr
[29] D.L. Wiltshire, \PR{D44} (1991) 1100.\cr
[30] G.W. Gibbons and K. Maeda, \NP{B298} (1988) 741.\cr
[31] M.S. Madsen and J.D. Barrow, \NP{B323} (1989) 242.\cr
[32] D.G. Boulware and S. Deser, \PRL{55} (1985) 2656.\cr
[33] D.L. Wiltshire, \PL{169B} (1986) 36; \PR{D38} (1988) 2445.\cr
[34] D.G. Boulware and S. Deser, \PL{173B} (1986) 409.\cr
[35] J.T. Wheeler, \NP{B273} (1986) 732;\br R.C. Myers and J.Z. Simon, \PR{D38}
(1988) 2434;\br B. Whitt, \PR{D38} (1988) 3000.\cr
[36] J. Hoell and W. Priester, Astron.\ Astrophys.\ {\bf 251} (1991) L23.\cr
[37] B. Carter, ``Mathematical foundations of the theory of relativistic
stellar and black hole configurations'', in {\it Gravitation in Astrophysics},
eds.\ B. Carter and J.B. Hartle, (Plenum Press, New York, 1987).\cr
[38] A. Papapetrou, \AIHP{A4} (1966) 83;\cr
[39] B. Carter, \JMP{10} (1969) 70.\cr
[40] B. Carter, \CMP{30} (1973) 261.\cr
}\bye